\documentclass[twocolumn,a4paper,prb,accepted=2026-08-17]{quantumarticle}
\pdfoutput=1

\usepackage[utf8]{inputenc}
\usepackage[T1]{fontenc}
\usepackage[english]{babel}
\usepackage{caption}
\usepackage{comment}
\usepackage{import}
\usepackage{enumerate}
\usepackage{amsmath}
\usepackage{amssymb}
\usepackage{amsthm}
\usepackage{mathtools}
\usepackage{stmaryrd}
\usepackage{graphicx}
\usepackage{centernot}
\usepackage{gensymb}
\usepackage{marginnote}
\usepackage{wasysym}
\usepackage{physics}
\usepackage{empheq}
\usepackage{tikz}
\usepackage{qcircuit}
\usepackage{color, soul}
\usepackage{xcolor}
\usepackage{verbatim}
\usepackage{enumitem}

\usepackage{framed}
\usepackage{empheq}
\usepackage{dsfont}

\definecolor{royalazure}{rgb}{0.0, 0.22, 0.86}
\usepackage[colorlinks=true,linkcolor=blue,citecolor=royalazure
]{hyperref}%
\usepackage{xurl}

\newcommand{\myappendixref}[1]{
    \hyperref[#1]{Appendix~\ref*{#1}}
}

\newtheorem{theorem}{Theorem}[section]

\theoremstyle{remark}

\newcommand{\bvecnodim}[1]{\ensuremath{\mathbf{e}_{#1}}}	

\newcommand{\hermset}[1]{\ensuremath{\mathcal{S}_{#1}}}		

\newcommand{\gstate}[1]{\ensuremath{\psi_0\left(#1\right)}} 
\newcommand{\estate}[2]{\ensuremath{\psi_{#2}\left(#1\right)}} 
\newcommand{\blambda}{\ensuremath{\boldsymbol{\lambda}}}    
\newcommand{\bdelta}{\ensuremath{\boldsymbol{\delta}}}      

\newcommand{\paulix}{\ensuremath{\sigma_x}}

\newcommand{\pauliz}{\ensuremath{\sigma_z}}

\DeclareMathOperator{\vecop}{\mathrm{vec}_r}		
\DeclareMathOperator{\traceop}{Tr}					
\DeclareMathOperator{\signf}{sign}					

\newcommand{\idenm}[1]{\ensuremath{\mathbb{I}_{#1}}}	
\newcommand{\idenmnodim}{\ensuremath{\mathbb{I}}}       

\usepackage{soul}

\newcommand{\newpart}[1]{\textcolor{black}{#1}}
\newcommand{\newnewpart}[1]{\textcolor{black}{#1}}

\begin{document}

\title{Order Parameter Discovery for Quantum Many-Body Systems}

\author{Nicola Mariella}
\email{nicola.mariella@ibm.com}
\affiliation{IBM Research - Dublin}
\author{Tara Murphy}
\email{tm763@cam.ac.uk}
\affiliation{IBM Research - Dublin}
\affiliation{Cavendish Laboratory, University of Cambridge,
J.J. Thomson Avenue, Cambridge CB3 0HE, United Kingdom}
\author{Francesco Di Marcantonio}
\email{francesco.di.marcantonio@cern.ch}
\affiliation{Department of Physical Chemistry and EHU Quantum Center, University of the Basque Country UPV/EHU, Box 644, 48080 Bilbao, Spain}
\author{Khadijeh Najafi}
\email{knajafi@ibm.com}
\affiliation{IBM Research, IBM T. J. Watson Research Center, Yorktown Heights, New York 10598, USA}
\author{Sofia Vallecorsa} 
\email{sofia.vallecorsa@cern.ch}
\affiliation{European Organization for Nuclear Research (CERN), Geneva 1211, Switzerland}
\author{Sergiy Zhuk}
\email{sergiy.zhuk@ie.ibm.com}
\affiliation{IBM Research - Dublin}
\author{Enrique Rico}
\email{enrique.rico.ortega@gmail.com}
\affiliation{Department of Physical Chemistry and EHU Quantum Center, University of the Basque Country UPV/EHU, Box 644, 48080 Bilbao, Spain}
\affiliation{European Organization for Nuclear Research (CERN), Geneva 1211, Switzerland}
\affiliation{Donostia International Physics Center, 20018 Donostia-San Sebastián, Spain}
\affiliation{IKERBASQUE, Basque Foundation for Science, Plaza Euskadi 5, 48009 Bilbao, Spain}

\begin{abstract}

Quantum phase transitions reveal deep insights into the behavior of many-body quantum systems, but identifying these transitions without prior knowledge of order parameters remains a significant challenge. In this work, we introduce a method for constructing phase diagrams using the vector field of the reduced fidelity susceptibility (RFS), and demonstrate how information encoded in this vector field can be used to discover observables corresponding to order parameters. We apply our approach to well-established models: the Axial Next Nearest Neighbour Interaction (ANNNI) model, a cluster state model, and a chain of Rydberg atoms; and validate the discovered order parameters using eigendecomposition and finite-size scaling analysis, confirming the expected universality classes. Our results demonstrate that the RFS vector field offers a unified framework for phase characterization and order-parameter discovery that requires no prior knowledge of symmetry or transition type, while relying only on reduced density matrices of small subsystems.
\end{abstract}

\maketitle
\section{Introduction}
\label{section:intoduction}

\begin{figure*}[ht]
\centering
\includegraphics[scale=0.3]{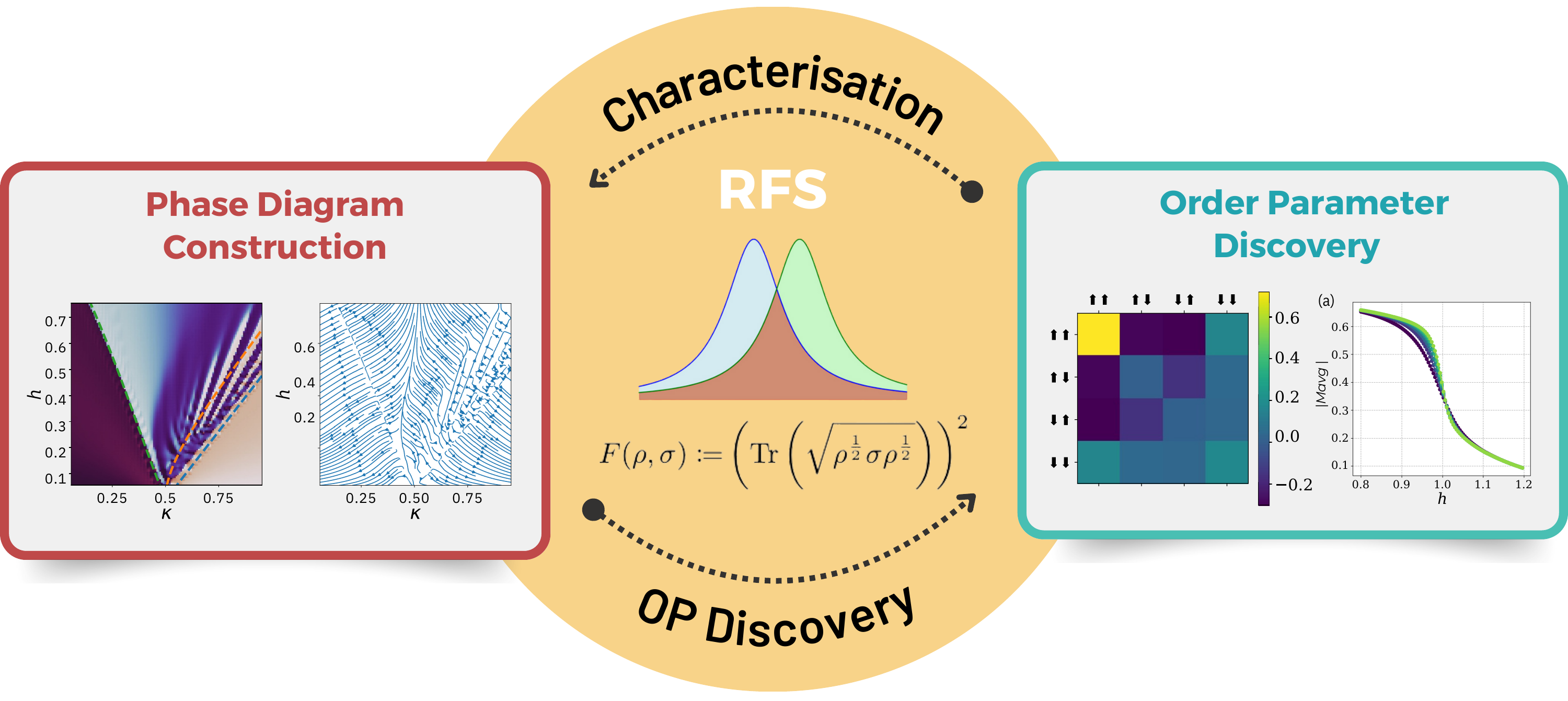}
\caption{\newpart{Visualisation of the reduced fidelity susceptibility (RFS) uses: phase diagram construction of quantum systems and order parameter discovery.}}
\label{fig:visual}
\end{figure*}

\newpart{Quantum phase transitions (QPTs) in many-body quantum systems pose considerable challenges for theoretical modeling and experimental observation. Unlike classical phase transitions, which are driven by thermal fluctuations, QPTs are induced by changes in external parameters such as magnetic fields or coupling constants, altering the ground-state properties of a system. The canonical framework for describing phase transitions is Landau–Ginzburg–Wilson (LGW) theory \cite{Ginzburg:1950sr, Sachdev_2011}, which classifies ordered phases by their broken symmetries and associated local order parameters. However, LGW theory famously fails for certain classes of transitions, most notably topological quantum phase transitions, where no local order parameter exists \cite{Hamma_2005, Levin_2006, Kitaev_2006}
}

\newnewpart{ 
Within the LGW paradigm, the Renormalization Group (RG) \cite{10.1119/1.11224, doi:10.1142/S0217751X13300500} and Finite-Size Scaling (FSS) theory \cite{ARDOUREL202399} provide the standard machinery for characterizing critical behavior: in formal terms, a QPT is defined only in the thermodynamic limit as a non-analyticity of the ground-state energy \cite{Sachdev_2011}, and FSS provides a systematic framework for extrapolating finite-size data to this limit. For transitions that fall outside the LGW classification, most notably topological transitions, where no local order parameter exists, complementary, order-parameter-agnostic diagnostics are needed. Machine learning techniques have been explored for this purpose \cite{Monaco_2023, cea2024exploring, PhysRevA.105.042432, PhysRevLett.132.207301, PhysRevX.12.031044}, as have fidelity-based probes, which we introduce next.}

\newpart{A particularly natural tool for detecting QPTs, which does not require a priori knowledge of the order parameter, is the quantum state fidelity \cite{jozsa-fidelity-for-mixed-states}. The fidelity susceptibility \cite{Dutta_Aeppli_Chakrabarti_Divakaran_Rosenbaum_Sen_2015, GU_2010, PhysRevE.76.022101} quantifies how sensitive a quantum state is to perturbations in the Hamiltonian, exhibiting sharp maxima near critical points where the ground state changes character most dramatically. Crucially, fidelity-based methods require no knowledge of the order parameter or the symmetry of the transition.}
\newnewpart{A parallel line of work replaces the global overlap with the fidelity between reduced density matrices (RDMs) of a subsystem. This \emph{reduced} (or partial-state) fidelity was connected to renormalization-group flows in Ref.~\cite{Zhou2007rgfidelity}; it was shown to signal level-crossing transitions for which the global fidelity is ill-behaved \cite{Kwok2008partial}; and exact results for small subsystems established both its distinct critical scaling, e.g., a $(\ln N)^2$ divergence of the two-site reduced fidelity susceptibility in the transverse-field Ising chain \cite{Ma2008rfsIsing}, an $N^{2/3}$ scaling in the Lipkin--Meshkov--Glick model \cite{Ma2008rfsLMG}, and a direct relation to the second derivative of the ground-state energy in chains with SU(2) and translation symmetry \cite{Xiong2009rfsHeisenberg}, and its practical appeal: experiments typically access subsystems rather than global states \cite{Ma2008rfsLMG}; see Section VI.B of the review \cite{GU_2010} for a comprehensive account. Mixed-state fidelities and Bures metrics have also been studied for thermal-state manifolds \cite{PhysRevA.75.032109, PhysRevA.76.062318}, a related but distinct setting. The present work builds on the reduced-fidelity tradition and extends it to two-dimensional parameter spaces, subsystems of arbitrary small size $k$, and the discovery of explicit order parameters.}

\newpart{Once a phase diagram is established, a separate but equally important challenge is the identification of an order parameter that characterizes each phase. \cite{PhysRevLett.96.047211} proposed a systematic approach: given representative RDMs from two phases, the observable that maximally distinguishes them is found by solving a simple optimization problem. Our work builds upon and extends this approach in two key ways: first, we use the RFS vector field to identify all phase boundaries simultaneously, without requiring prior selection of representative states; second, we generalize the pairwise optimization of \cite{PhysRevLett.96.047211} to simultaneously handle all pairs of phase labels identified from the vector field, allowing multi-phase order parameters to be discovered in a single optimization step \cite{PRXQuantum.2.010102}. A schematic picture of our method can be seen in \autoref{fig:visual}.}

\newpart{We describe the contributions of the present work. We construct a vector field from the RFS defined on a two-dimensional parameter space, and show that this vector field encodes the full phase diagram: sources correspond to phase transitions and sinks to the representative ground states of each phase. We then formulate an optimization problem that discovers Hermitian observables functioning as order parameters, using the phase labels extracted from the vector field as supervision. We demonstrate the efficacy of the method on three models: the ANNNI model \cite{PhysRevLett.93.056402, Selke1988TheAM}, the cluster Hamiltonian, and a Rydberg atom chain. We provide a detailed analysis of the Rydberg ripple features, supporting their physical origin as incommensurate phase boundaries \cite{PhysRevResearch.3.023049, PhysRevB.106.165124, PhysRevA.98.023614}.}

\section{Preliminary}
\label{sec:prelim}
We begin by defining the set of Hermitian matrices of order $n$, which is denoted by $\hermset{n} \coloneqq \{M \in \mathbb{C}^{n\times n}|M=M^{\dagger}\}$.
We denote the identity matrix of order $n$ by $\idenm{n}$ and the $i$th canonical basis vector for the vector space $\mathbb{C}^n$ by $\bvecnodim{i}$.
    
Next, we introduce the concept of fidelity. Fidelity measures the similarity between two quantum states, and all its definitions embody the probability of distinguishing one quantum state from another.
    
Given mixed-states $\rho, \sigma$ we define the \textit{Uhlmann-Jozsa fidelity} \cite{jozsa-fidelity-for-mixed-states} by\footnote{Some literature defines the fidelity in \eqref{eq:uhlmann-fidelity-def} as the square root of our $F$ (e.g. \cite{Nielsen}).}
\begin{align}
\label{eq:uhlmann-fidelity-def}
F(\rho, \sigma)\coloneqq&\left(\traceop\left(\sqrt{\rho^{\frac12}\sigma\rho^{\frac12}}\right)\right)^2,
\end{align}
which, for the remainder of the work, will be referred to as \textit{fidelity}, unless otherwise specified. The latter definition of fidelity is not unique, indeed, Jozsa \cite{jozsa-fidelity-for-mixed-states} obtained an axiomatic definition for a family of such functions. However, the Uhlmann fidelity stands for its unique relation to the \textit{Bures distance}
\begin{align}
\label{eq:sq-bures-dist-def}
d_B^2(\rho, \sigma) \coloneqq 2 \left(1 - \sqrt{F(\rho, \sigma)}\right),
\end{align}
which is a quantification of the statistical distance between density matrices. A relevant result is the \textit{Uhlmann theorem} \cite{Uhlmann1976TheP}, which expresses the fidelity in terms of the maximum overlap over all purifications of its arguments, that is
\begin{align}
\label{eq:uhlmann-thm}
F(\rho, \sigma) =& \underset{\ket{\psi_\rho}}{\max} \left|\bra{\psi_{\rho}}\ket{\psi_{\sigma}}\right|^2,
\end{align}
where $\ket{\psi_{\sigma}}$ and $\ket{\psi_{\rho}}$ denote some purifications of $\sigma$ and $\rho$, respectively.

Following our discussion on the fidelity, we next consider a generic many-body parametric Hamiltonian $H(\lambda)=H_0 + \lambda H_I$, with eigenvalue equation $H(\lambda)\ket{\estate{\lambda}{k}}=E_k(\lambda) \ket{\estate{\lambda}{k}}$, where the index $k=0$ identifies the ground state.
For clarity, we may omit the dependence of the various objects from the parameter $\lambda$, when it does not result in ambiguity.
In a perturbative approach w.r.t. a small parameter $\delta$, the Taylor expansion for the fidelity reads
\begin{align}
\label{eq:fidelity-sus-def}
\left|\bra{\gstate{\lambda}}\ket{\gstate{\lambda+\delta}}\right| =& 1 - \frac{1}{2} \mathcal{X}_F(\lambda) \delta^2 + O\left(\delta^3\right),
\end{align}
with the leading term (second-order) revealing the \textit{fidelity susceptibility} $\mathcal{X}_F$ (susceptibility for short) \cite{PhysRevE.76.022101}. We note that the first-order term vanishes since the fidelity reaches a maximum at $\delta=0$ for any $\lambda$. Informally, fidelity susceptibility quantifies how much the fidelity (or overlap) between two nearby states changes concerning a small change in a parameter (such as an external field or coupling strength) that defines the states. Let $\rho_0(\lambda)=\ket{\gstate{\lambda}}\bra{\gstate{\lambda}}$ be the density matrix for the ground state of $H(\lambda)$. Two well-known formulations for the susceptibility\footnote{The rightmost of \eqref{eq:fidelity-sus-def-ii} is obtained by considering the Taylor expansion of $\ln(1+x)=x-\frac{x^2}{2} + O(x^3)$, with $x=- \frac{1}{2} \mathcal{X}_F(\lambda) \delta^2 + O\left(\delta^3\right)$ from \eqref{eq:fidelity-sus-def}.} are given as;
\begin{subequations}
\begin{align}
\label{eq:fidelity-sus-def-ii}
\mathcal{X}_F(\lambda) =& -\frac{\partial^2 \sqrt{F(\rho_0(\lambda), \rho_0(\lambda + \delta))}}{\partial \delta^2}\Big|_{\delta=0}\\
=& \underset{\delta \to 0}{\lim} -\frac{2\ln \sqrt{F(\rho_0(\lambda), \rho_0(\lambda + \delta))}}{\delta^2},
\end{align}
\end{subequations}
where $F$ is generalized to the Uhlmann fidelity in \eqref{eq:uhlmann-fidelity-def}, when the arguments are not pure states.
The expansion for the ground state of the perturbed Hermitian\footnote{This result is commonly known as the Rayleigh–Schr\"{o}dinger perturbation theory \cite{Schrodinger-perturb}.} $H(\lambda)$ is given as;
\begin{align}
\ket{\gstate{\lambda + \delta}} \approx&  \ket{\gstate{\lambda}} + \delta
\sum_{k\ne 0} \frac{
K_{k, 0}
}{E_k - E_0}
\ket{\estate{\lambda}{k}},
\end{align}
where $K_{k, 0}\coloneqq \bra{\estate{\lambda}{k}}H_I\ket{\gstate{\lambda}}$. This leads to another well-known formulation for the susceptibility
\begin{align}
\label{eq:susceptibility-from-H-spectrum}
\mathcal{X}_F(\lambda) =&
\sum_{k\ne 0} \frac{
\left|
\bra{\estate{\lambda}{k}}H_I\ket{\gstate{\lambda}}
\right|^2
}{(E_k - E_0)^2} \ge 0.
\end{align}
As noticed in \cite{Wang_2015}, the latter, which depends solely on the spectrum, shows that the quantity $\mathcal{X}_F$ is non-negative and diverges when the energy gap closes.

\section{Phase Diagram Construction}
\label{section:main}
Building upon the previous section's exploration of fidelity susceptibility, we next introduce the reduced fidelity susceptibility (RFS) vector field concept. In doing so, we consider a general many-body Hamiltonian parameterized by the space $\mathcal{X} \subseteq \mathbb{R}^2$, of which its decomposition in terms of base and driving components reads
\begin{align}
    H(\blambda) =& H_0 + \lambda_1 H_1 + \lambda_2 H_2,
    \label{eq:hamil}
\end{align}
with control parameters $\left(\lambda_1\quad\lambda_2\right)^\top \in \mathcal{X}$.
We anticipate that choosing the two-dimensional parameter space is convenient for the visual approach, but not fundamental. Let $\ket{\gstate{\blambda}} \in \mathcal{H}$ denote the ground state of the Hamiltonian $H$ evaluated at $\blambda \in \mathcal{X}$, with $\mathcal{H}$ denoting the underlying Hilbert space. The Hilbert space $\mathcal{H}$ is assumed bipartite, that is $\mathcal{H}=\mathcal{H}_\mathcal{A} \otimes \mathcal{H}_\mathcal{B}$, for some subsystems $A$ and $B$. In addition, we assume the ground state is non-degenerate $\psi_0$.

\newpart{Choice of bipartition. The RFS is defined relative to a bipartition $\mathcal{H}=\mathcal{H}_\mathcal{A} \otimes \mathcal{H}_\mathcal{B}$ of the full Hilbert space, where $\rho_0(\blambda) = Tr_\mathcal{B}(\psi_0 \psi_0^*) $ is the RDM of subsystem $\mathcal{A}$. In practice, we choose $A$ to consist of $k$ contiguous central sites of the chain, and the results are robust to moderate changes in the location of $\mathcal{A}$, provided $\mathcal{A}$ is in the bulk. The choice of $k$ is more important; we propose the following practical criterion: increase $k$ until the phase diagram stabilizes. For the models studied here, $k = 1$ or $2$ suffices to reconstruct the phase diagram, except for the cluster Hamiltonian, which requires $k = 5$ to capture the non-local string order. As a fail-safe, the indefiniteness condition on the order-parameter solution (see Appendix \ref{section:ord-param-solution}) signals when $k$ is insufficient for order-parameter discovery.}

Let $\rho_0(\blambda)$ denote the reduced density matrix (RDM) resulting from tracing out the subsystem $B$ for the ground state, so
\begin{align}
\label{eq:gstate-rho}
\rho_0(\blambda)=\traceop_B\left(
\ket{\gstate{\blambda}}
\bra{\gstate{\blambda}}
\right).
\end{align}

We consider a perturbative action in the parameter space and define
\begin{align}
\label{eq:fun-f-def}
f(\boldsymbol{\lambda}, \boldsymbol{\delta}) =& \sqrt{F\left(\rho_0(\boldsymbol{\lambda}), \rho_0(\boldsymbol{\lambda} + \bdelta)\right)}
\in [0, 1],
\end{align}
for $\blambda \in \mathcal{X}$ and $\bdelta$ a perturbation in the latter space, where $F$ is the fidelity defined in \eqref{eq:uhlmann-fidelity-def}.
As a consequence of the Uhlmann theorem \eqref{eq:uhlmann-thm}, since the fidelity is the maximum over the overlap w.r.t. all purifications, then
$f(\boldsymbol{\lambda}, \boldsymbol{\delta}) \ge \left|
\bra{\gstate{\blambda}}
\ket{\gstate{\blambda + \bdelta}}
\right|$,
that is, the quantity in \eqref{eq:fun-f-def} is bounded below\footnote{Alternatively, this can be inferred by casting the partial trace as Kraus operator and by the monotonicity of the Uhlmann Fidelity \cite{Mendon_a_2008}.}
by the square root of the overlap between the perturbed ground states.

Following this, we introduce one of the key functions for our method\footnote{
    Interestingly, we can obtain another function $g$ proportional to that in \eqref{eq:fun-g-def} by considering $f$ defined in terms of the squared Bures distance \eqref{eq:sq-bures-dist-def}.
}, that is
\begin{align}
\label{eq:fun-g-def}
g(\boldsymbol{\lambda}) \coloneqq& -\left(\frac{\partial^2 f(\blambda, \bdelta)}{\partial \delta_1^2}
+ \frac{\partial^2 f(\blambda, \bdelta)}{\partial \delta_2^2} \right)\biggr\rvert_{\bdelta=\mathbf{0}}.
\end{align}
We call the terms $\partial^2 f/\partial \delta_k^2$ the \textit{reduced fidelity susceptibility} for the corresponding parameter $\lambda_k$.
Indeed, it can be noted that the terms of the latter take the form of the fidelity susceptibility in \eqref{eq:fidelity-sus-def-ii}.
Moreover, as shown in \eqref{eq:susceptibility-from-H-spectrum}, the susceptibility is a non-negative quantity, then it is reasonable to consider a quantity resulting
from the summation of susceptibilities for different directions (parameters).
The adjective reduced is justified by the fact that we are considering RDMs instead of pure states as in \eqref{eq:fidelity-sus-def}. 

\newnewpart{Susceptibilities of reduced states have been studied before: for ground-state RDMs, general formulas and exact critical scaling were obtained for small subsystems \cite{Zhou2007rgfidelity, Kwok2008partial, Ma2008rfsLMG, Ma2008rfsIsing, Xiong2009rfsHeisenberg} (see also Section VI.B of \cite{GU_2010}), while related Bures-metric constructions were developed for thermal-state manifolds \cite{PhysRevA.75.032109, PhysRevA.76.062318}. Our $g(\blambda)$ belongs to the first setting: it is the trace of the Bures metric tensor of the ground-state RDM manifold (\myappendixref{section:info-geo-link}).}

We also note that for the regular susceptibility, it is clear that the linear term in the perturbative Taylor expansion of $f$ vanishes for $\bdelta=\mathbf{0}$.
In \myappendixref{section:info-geo-link} we expand on some key concepts in information geometry and its connection to the definition in \eqref{eq:fun-g-def}.

We next obtain the vector field $P: \mathcal{X} \to \mathbb{R}^2$ (under sufficient smoothing assumptions) defined by;
\begin{align}
\label{eq:vfield-def}
P(\blambda)\coloneqq& -\nabla_{\boldsymbol{\lambda}}g(\blambda).
\end{align}    
We stress that the gradient in \eqref{eq:vfield-def} is expressed w.r.t. $\boldsymbol{\lambda}$ (parameters vector), whereas the Laplacian in \eqref{eq:fun-g-def} is related to $\bdelta$ (perturbation).
In addition, we obtain a scalar function mapping the parameters $\blambda$ to the angles of the vectors in the image of $P$, that is
\begin{align}
\label{eq:angle-grad}
\theta(\blambda) =& \mathrm{Arg}\left(
\bvecnodim{1}^\top P(\blambda) + \imath \bvecnodim{2}^\top P(\blambda)
\right),
\end{align}
with $\mathrm{Arg}: \mathbb{C} \to (-\pi, \pi]$ denoting the \textit{principal argument}\footnote{
    Let $z=x+\imath y$ with $x, y \in \mathbb{R}$, then the principal argument of $z$ is defined by
    $\mathrm{Arg}(x+\imath y)=\mathrm{atan2}(y, x)$,
    with $\mathrm{atan2}(y, x)=\lim_{c \to x^+}\arctan\left(\frac{y}{c}\right) + \frac{\pi}{2}\signf(y)\signf(x)\left(\signf(x)-1\right)$.
}.
The latter is defined on the subset of the parameter space $\mathcal{X}^{\prime}=\left\{\blambda \in \mathcal{X}\middle| P(\blambda)\ne\mathbf{0}\right\}$.
We note, as outlined in \myappendixref{section:scale-invariant}, the function in \eqref{eq:angle-grad} is scale invariant in the proximity of QPT at $\blambda_c$,
that is $\theta((\blambda - \blambda_c)t)=\theta(\blambda - \blambda_c)$ for some small $t>0$.
The latter consideration is justified by the scaling behavior of fidelity susceptibility in the vicinity of QPT \cite{PhysRevB.81.064418}.

\begin{figure}
\centering
\includegraphics[scale=0.23]{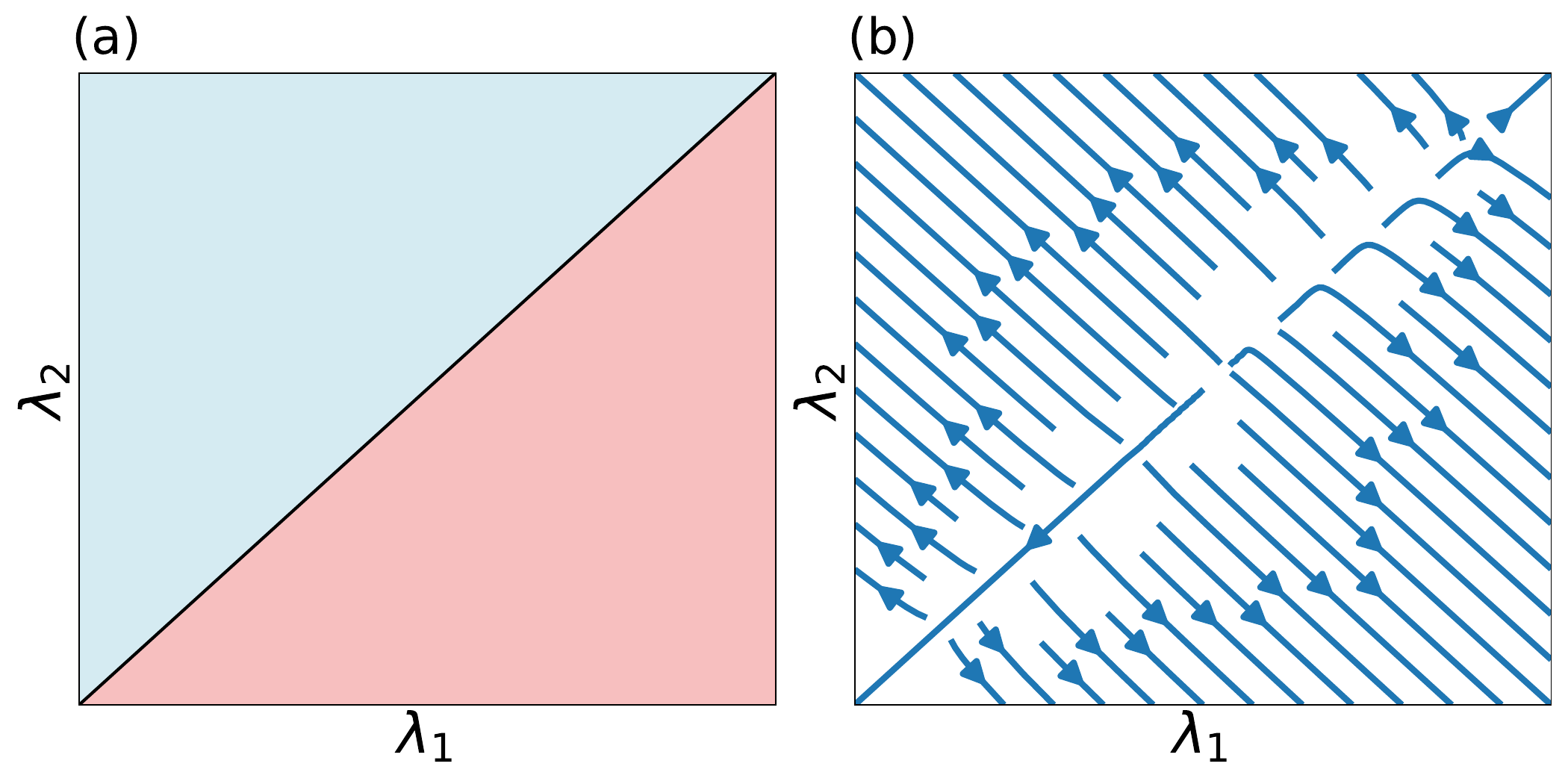}
\caption{(a) Color-mapped visualization of the fidelity vector field angle $\theta(\blambda)$.
(b) Quiver plot of the vector field $P(\blambda)$ at a phase transition line.
We note that the angle (thus the associated color) drastically changes at the phase transition in (a),
which corresponds to a source (arrows radiate outward) in the vector field in (b). 
}
\label{fig:sketch2}
\end{figure}

The function $g$ defined in \eqref{eq:fun-g-def} corresponds to a notion of fidelity susceptibility. Moreover, the angle given by $\theta(\blambda)$ in \eqref{eq:angle-grad} is the direction of the steepest decline in susceptibility.
Consequently, we expect that phase transitions materialize as sources in the vector field \eqref{eq:vfield-def}, that is, the loci where the susceptibility is maximized.
An example of this can be seen in \autoref{fig:sketch2}. Furthermore, we note that the sinks are the points where the susceptibility reaches the local minimum. As the last step, if we map the co-domain of $\theta:\mathcal{X}^{\prime} \to (-\pi, \pi]$ to a cyclic color map, we expect to obtain the phase diagram of the Hamiltonian in \eqref{eq:hamil}.
This process is depicted in the sketch shown in \autoref{fig:sketch2}, where in panel (a) we plot the angle $\theta(\blambda)$, and panel (b) represents the vector field with phase transition depicted by a diagonal line.
Both panels suggest that in the proximity of a QPT, the gradient in \eqref{eq:vfield-def} changes heading abruptly.
We elaborate on this in \myappendixref{section:vfield-patterns}.
The practical implementation of the construction is expanded in \myappendixref{section:practical-phase-diag-construction}.

\newpart{We conclude with a remark on extensions. Constant-size RDMs are particularly effective for phases characterized by local order parameters or short-range entanglement. For phases with topological order, where no local order parameter exists, the relevant information is encoded in the global entanglement structure, and larger subsystems (capturing string correlations) or nonlinear functionals of the RDM (such as entanglement entropy or topological entanglement entropy) must be used. The present paper focuses on the case where $k$ is kept small but demonstrates, through the cluster Hamiltonian example, that the method naturally adapts to non-local order by increasing $k$.}

\section{Order parameter discovery}
\label{section:ord-param}
Following the results obtained in the preceding section, we next investigate whether it is possible to determine an order parameter that captures the phase transitions of a given Hamiltonian.
Specifically, we consider the class of order parameters that are determined by the expectation of a Hermitian observable.

To answer this question, we apply the protocol as described in \autoref{section:main}, and move from susceptibility-based identification of phases to determining the optimal observable
for distinguishing the phases, that is, a \textit{local order parameter}.
A simpler method has been proposed in \cite{PhysRevLett.96.047211}, where a Hermitian operator is optimized to distinguish two given RDMs. In the latter, the distinction is given by the sign of the expectation with respect to the optimal observable. Our contribution, instead, aims at mimicking the behaviour of order parameters.

We first note that the function $\theta(\blambda) \in (-\pi, \pi]$ in \eqref{eq:angle-grad} represents the point-wise angle of the vector field \eqref{eq:vfield-def}. Also, as argued before, phase transitions are likely to materialize as sources, and we assume that adjacent phases are distinguished by $\pi$ radians. In other words, if the parameter $\blambda_c \in \mathcal{X}$ is proximal to a critical point (for some Hamiltonian),
and $\bdelta$ a perturbation in the same space, then the assumption is that the corresponding gradient vectors have opposite directions, that is $\left|\theta(\blambda_c)-\theta(\blambda_c + \bdelta)\right| \approx \pi$. Consequently, there exists an optimal angle $\eta$ such that
\begin{align}
    \label{eq:sin-p-opposite-sign}
    \left|
        \sin\left(\theta(\blambda_c) + \eta\right) +
        \sin\left(\theta(\blambda_c  + \bdelta) + \eta\right)
    \right| \approx 0.
\end{align}

Consider a parametric Hamiltonian on $n$ spins with ground state RDM $\rho_0(\blambda)$ defined as in \eqref{eq:gstate-rho}. 
In addition, we assume that the RDM is small enough to be handled classically.
Given a finite set of parameters $\{\blambda_i\}$ in the neighbour of $\blambda_c$,
we define a label $y_i \in [-1, 1]$ for each parameter $\blambda_i$ as $y_i=\sin\left(\theta(\blambda_i) + \eta\right)$. From \eqref{eq:sin-p-opposite-sign}, we see that distinct phases will be assigned opposite signs, $\mathrm{sign}(y_i)$. We define the index sets $I^+=\{i|y_i>0\}$ and $I^-=\{i|y_i<0\}$, partitioning the indices for the parameters $\{\blambda_i\}$ determining the ground states laying on the ordered and disordered phases, respectively.
Under the assumption of non-degeneracy of the problem (see \myappendixref{section:ord-param-solution} for more details),
we devise the following (non-convex)
\textit{quadratically constrained quadratic program} (QCQP) \cite{Boyd_Vandenberghe_2004},
\begin{align}
    \label{eq:ord-param-qcqp}
    \nonumber
    \underset{M \in \hermset{m}}{\min}&
    \sum_{(i, j) \in I^+ \times I^-} \left(
    -\frac{\langle M \rangle_i^2}{p_i}
    +
    \frac{\langle M \rangle_j^2}{p_j}
    \right),\\
    \text{s.t.}\,& \|M\|_F^2 \le 1,
\end{align}
with $\langle M \rangle_i\coloneqq\traceop(\rho_0(\blambda_i)M)$ defining the expectation of $M$ at $\blambda_i$,
and $p_i\coloneqq \traceop(\rho_0(\blambda_i)^2)$ the purity of the same RDM.
We denoted by $\hermset{m}$ the set of Hermitian matrices of order $m$, which is also the order of the RDM $\rho_0(\blambda_i)$.

This is commonly known in optimization literature as the \textit{trust region problem} \cite{trust-region-methods}, which can be interpreted as a generalization of the minimum eigenvalue problem. The problem is solvable efficiently (polynomial time) even in the cases where the quadratic term is not positive semidefinite (i.e. non-convex) \cite{Rendl1997}.

\newpart{The problem in Eq. \eqref{eq:ord-param-qcqp} is solvable efficiently in polynomial time in the dimension of $M$ \cite{Rendl1997}. We note that for a $k$-site RDM in a spin-$1/2$ chain, the matrix $M$ has dimension $2^k \times 2^k$, so the storing cost scales as $O(4^k)$ per optimization, which is completely tractable for the values $k = 1, 2, 4, 5$ used in this work.}

Informally, the optimization favors the non-zero expectation $\langle M \rangle_i$ for the labels $y_i > 0$ (ordered phase), whereas the quadratic term $\langle M \rangle_i^2$ penalizes non-zero expectations for the labels $y_i < 0$ (disordered phase).
In essence, this mechanism is mimicking the order/disorder behavior of the order parameters, for all pairs of opposing (w.r.t. phase) RDMs $(\rho_0(\blambda_i), \rho_0(\blambda_j))$,
with $(i, j) \in I^+ \times I^-$.

\newcommand{\rhoa}{\ensuremath{\rho_+}}
\newcommand{\rhob}{\ensuremath{\rho_-}}
A special case of the problem in \eqref{eq:ord-param-qcqp} arises when we consider the reference density matrices $\rhoa$ and $\rhob$, each representing a phase (i.e. ordered vs disordered).
The formulation becomes
\begin{align}
    \label{eq:ord-param-special}
    \nonumber
    \underset{M \in \hermset{m}}{\min}&
    -\frac{\traceop\left(\rhoa M\right)^2}{\traceop\left(\rhoa^2\right)}
    +
    \frac{\traceop\left(\rhob M\right)^2}{\traceop\left(\rhob^2\right)},\\
    \text{s.t.}\,& \|M\|_F^2 \le 1.
\end{align}
In \myappendixref{section:ord-param-solution-special}, it is shown that the objective of the latter is the quadratic form of an operator connected to the 
\textit{Gram-Schmidt process} (GS) \cite{golub}, a method related to the \textit{$QR$ decomposition} in linear algebra.
For a density matrix $\rho$, we define $\widehat{\rho}\coloneqq \rho/\|\rho\|_F$ its associated norm-1 PSD
(positive semidefinite\footnote{
    That is, a Hermitian matrix with non-negative eigenvalues.
}).
Also, for Hermitian matrices $A, B$ of the same order, we denote by $\langle A, B\rangle$ their Frobenius inner product.
The special formulation leads to the close form solution
\begin{align}
    \label{eq:ord-param-special-em}
    M =& \frac{\widehat{\rhoa}-\left\langle \widehat{\rhoa}, \widehat{\rhob} \right\rangle \widehat{\rhob}}{
    \sqrt{1-\left\langle \widehat{\rhoa}, \widehat{\rhob} \right\rangle^2}},
\end{align}
when $\rhoa \ne \rhob$.
The solution is Hermitian as it is a real-linear combination of PSDs.
One can immediately verify that $\traceop(\rhob M)=0$, whereas
\begin{align}
    \traceop(\rhoa M) =&
    \|\rhoa\|_F \sqrt{1-\left\langle \widehat{\rhoa}, \widehat{\rhob} \right\rangle^2} >0,
\end{align}
as expected.
Geometrically, the solution in \eqref{eq:ord-param-special-em} can be interpreted as subtracting from $\widehat{\rhoa}$ its projection on
$\widehat{\rhob}$, hence the resulting Hermitian is orthogonal to $\rhob$ and has a non-zero overlap with $\rhoa$ (when $\rhoa\ne \rhob$).
Moreover, the latter provides an understanding of the general problem.
Specifically, the objective in \eqref{eq:ord-param-qcqp} corresponds to the sum of special objectives in \eqref{eq:ord-param-special},
for all pairs of opposing indices $(i, j) \in I^+ \times I^-$.

In practical terms, we consider RDMs on a few sites, thus the quadratic optimization problem can be solved classically upon computation of the partial trace on the ground states. Experimental results are presented in \autoref{section:experiments-ord-params} and the details for the solution of the optimization are expanded in \myappendixref{section:ord-param-solution}.

We remark, as discussed in \cite{goldenfeld2018lectures}, that the order parameter need not be unique, and any scaling operator that is zero in the disordered phase and non-zero in an adjacent (on the phase diagram), usually ordered phase, is a possible choice for an order parameter.


\section{Experiments}
\label{section:experiments}
To demonstrate the potential of the fidelity vector field in identifying QPTs and their corresponding order parameters, we apply our above methods to the cluster Hamiltonian and ANNNI models. 

\subsection{The ANNNI model}
\label{section:annni}

We first focus on the ANNNI Model \cite{PhysRev.124.346, PhysRevLett.44.1502, Selke1988TheAM}, a theoretical model commonly used in condensed matter physics to study the behavior of magnetic systems. 
The Hamiltonian of such a system can be written as 
\begin{align}
\label{eq:annni}
H =& -J_1\sum_{i=1}^{N-1}\sigma^{x}_{i}\sigma^{x}_{i+1}-J_{2}\sum_{i=1}^{N-2}\sigma^{x}_{i}\sigma^{x}_{i+2}-B\sum_{i=1}^{N}\sigma^{z}_{i},
\end{align}
which we can rewrite in terms of the dimensionless ratios $\kappa=-J_2/J_1$ and $h=B/J_1$. The former is called the \textit{frustration parameter} while the latter is related to the transverse magnetic field. 
With respect to the general Hamiltonian in \eqref{eq:hamil}, the parameters $\lambda_1, \lambda_2$ are now associated respectively to the parameters $\kappa, h$ in \eqref{eq:annni}.

In this model, spins are arranged in a one-dimensional lattice, with each spin existing in either the $\ket{\uparrow}$ or $\ket{\downarrow}$ state. The inclusion of the nearest neighbor and next-nearest-neighbor interactions in this model introduces a type of frustration, where the optimal alignment of neighboring spins is hindered due to competing interactions. The transverse field present, which represents an external magnetic field perpendicular to the direction of the spins, also induces quantum effects and modifies the overall behavior of the system. It is the combination of these complex interactions, which combine the effect of quantum fluctuations (owing to the presence of a transverse magnetic field) and frustrated exchange interactions that lead to phenomena such as quantum phase transitions. As a consequence, it is a paradigm for the study of competition between magnetic ordering, frustration, and thermal disordering effects \cite{2007JPhA...40.6251C}.

\begin{figure}[h!]
\centering
\includegraphics[scale=0.43]{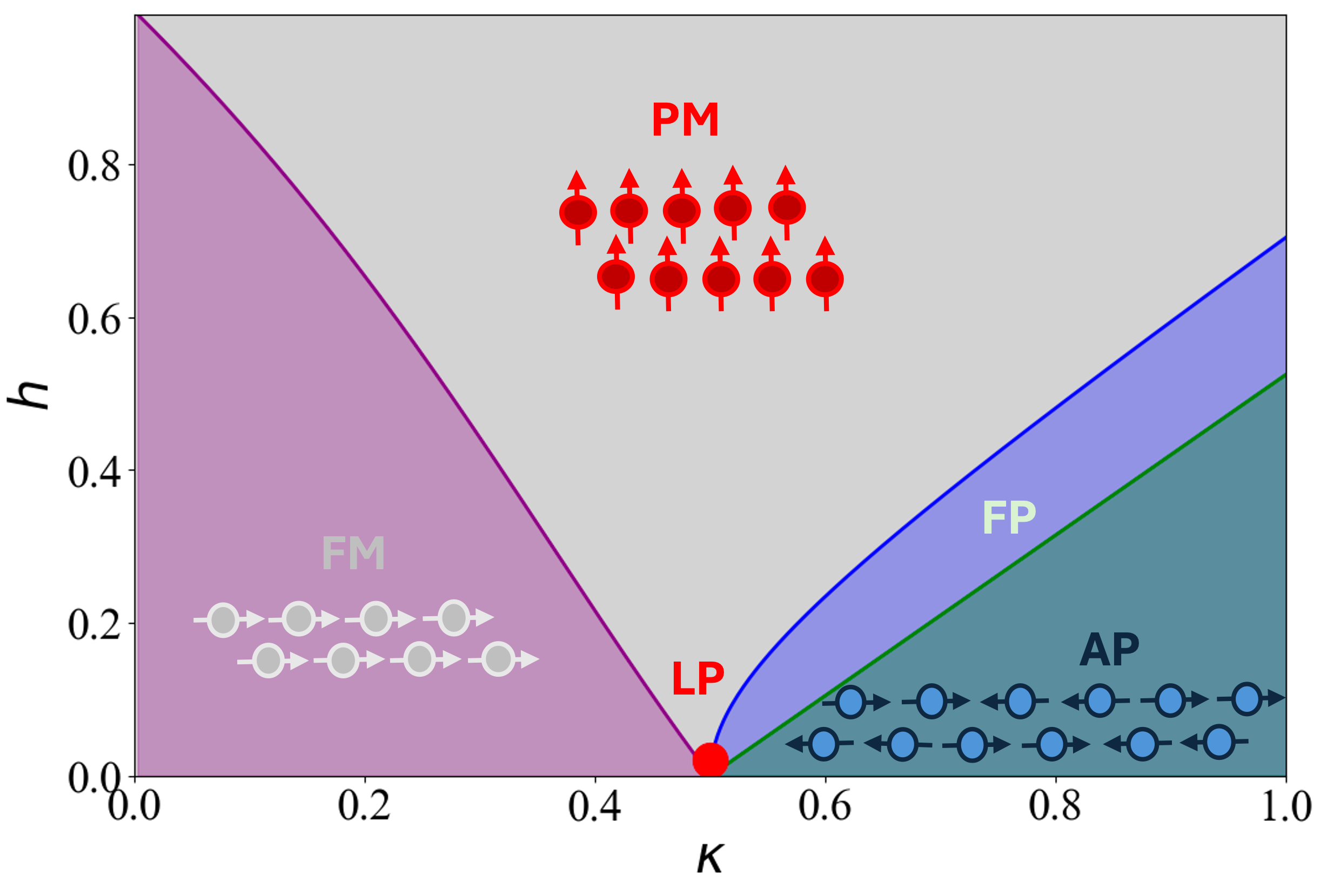}
\caption{(a) Phase Diagram of one-dimensional ANNNI Model. Here the pink represents the Ferromagnetic Phase (FM), where all spins are aligned along the x direction. The grey area corresponds to the paramagnetic phase (PM), in which the magnetic field dominates and all spins align along the z direction. The antiphase (AP) corresponds to the green region where the ground state takes the form of a staggered magnetization pattern with period four. Both the PM and AP are separated by the floating phase (FP). Here the spin chain can be seen as a ladder of two spin chains
as sketched in the cartoon spin configurations.}
\label{fig:annni-theory}
\end{figure}

The phase diagram can be seen in \autoref{fig:annni-theory}, where three phase transitions are present. The first is the Ising-like transition between the Ferromagnetic (FM) and Paramagnetic (PM) phases, the second is the Kosterlitz-Thouless (KT) transition between the paramagnetic and floating phase (FP), and the final is the Pokrovsky-Talapov (PT) phase transition between the floating phase and antiphase (AP).

For more information on each of these phases and the corresponding phase transitions and their ground states see \myappendixref{sec:annnitheory}.

\begin{figure*}[ht]
\centering
\includegraphics[scale=0.5]{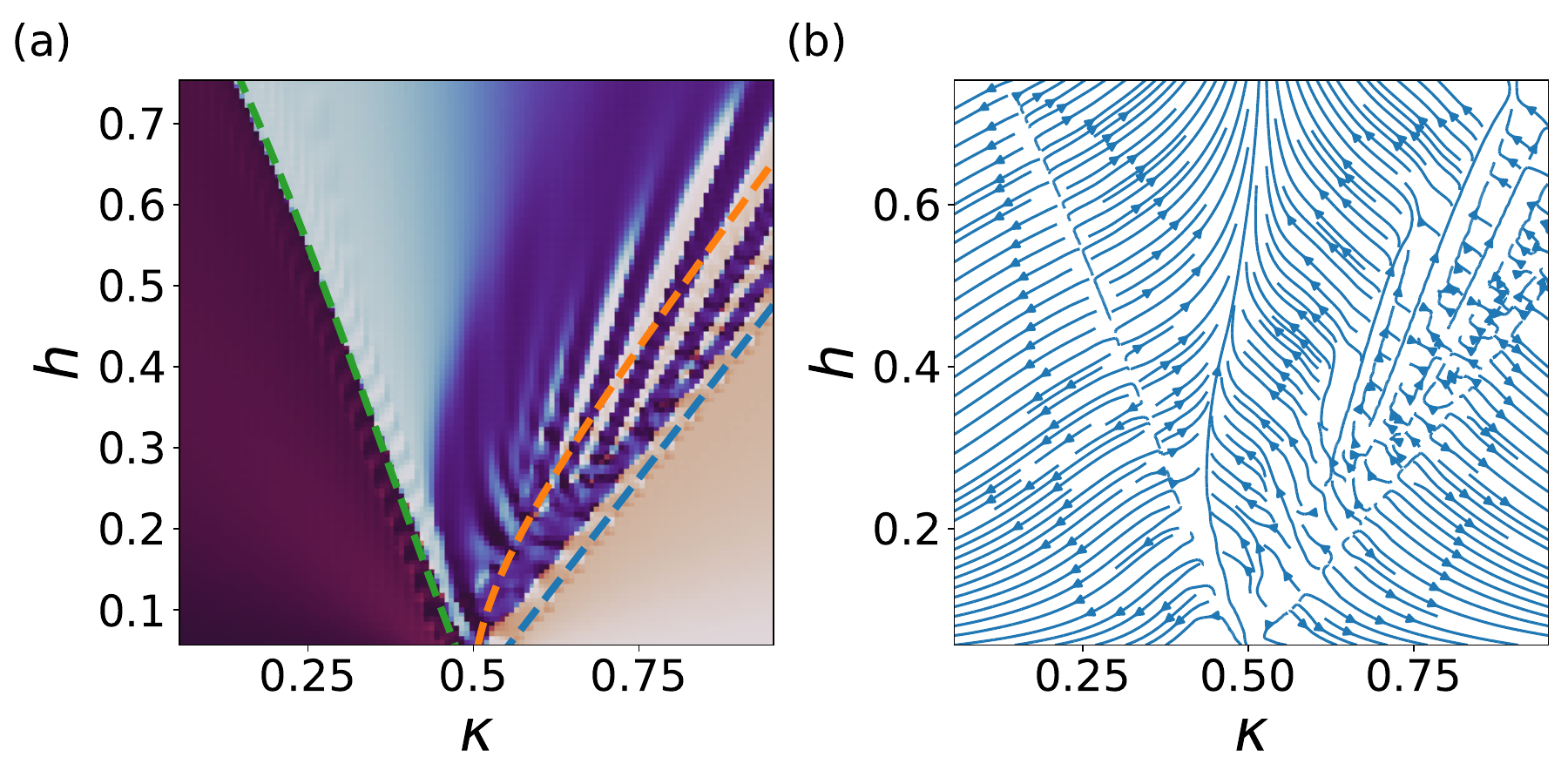}
\caption{Phase Diagram obtained using the reduced fidelity susceptibility of the one-dimensional ANNNI Model. \newnewpart{Dashed lines indicate the known Ising, KT, and PT phase boundaries reported in \myappendixref{sec:annnitheory}.} Here a chain length of $L = 50$ spin sites was used and a \newnewpart{two-site RDM} was used when calculating the gradient of the reduced fidelity susceptibility. (a) The angle of the vector field given in \eqref{eq:angle-grad} is plotted  (b) the vector field given in \eqref{eq:vfield-def} is plotted.}
\label{fig:phasesd-annni-i}
\end{figure*}

Using the formalism as outlined in \autoref{section:main}, we first obtain the ground states of the Hamiltonian using the density matrix renormalization group (DMRG).
Next, we consider a two-site RDM and calculate the RFS vector field and its corresponding angles. For details on how the RDM was obtained or the DMRG algorithm,
see \myappendixref{section:dmrg_computation}.
In \autoref{fig:phasesd-annni-i} (a), we plot the color-mapped angle as given in (\ref{eq:angle-grad}),
as well as the numerical estimation for the Ising, KT, and PT phase transition lines,
obtained in \cite{PhysRevB.76.094410, cea2024exploring, Suzuki2013} (additional details given in \myappendixref{sec:annnitheory}).
There is a noticeable overlap between our approach and the literature.
However, we cannot expect a perfect overlap, because of the finite-size scaling effects.
As predicted in the method formulation, we have numerical confirmation from the vector field plotted in \autoref{fig:phasesd-annni-i} (b),
that sources indeed correspond to phase transitions.
This source and sink-like behavior, as seen in \autoref{fig:phasesd-annni-i} (b) corresponds to local maxima and minima of the RFS in \eqref{eq:fun-g-def}
and is further analyzed in \myappendixref{section:vfield-patterns}.

Additionally, it is worth noting that the fidelity vector field surpasses previous simulations in its ability to capture intricate details that were previously unresolved in the floating phase \cite{PhysRevE.75.021105}. Distinct ripples-like features are present, with the sources of the vector field corresponding to each of these structures indicating the presence of the phase transition between the paramagnetic and floating phases, specifically the KT transition.
Future work will aim to explore the structure and fully utilize the potential of the RFS vector field in analyzing such complex phenomena.
In the subsequent section, we demonstrate how we apply this method to determine order parameters for a given quantum system. 

\subsection{Order parameters discovery}
\label{section:experiments-ord-params}
Next, we experimentally demonstrate the validity of the method used for order parameter discovery (\autoref{section:ord-param}), and in doing so, propose a method for understanding the structure of the optimal observable. We proceed with the ANNNI model \eqref{eq:annni} by considering the phase diagram in the region $(\kappa\quad h) \in \mathcal{R}=[0.5, 2.1] \times [0, 1.6]$ and begin with the use of the RDMs of single spin sites.
In other words, we focus on the side of the diagram (see \autoref{fig:annni-theory}) that lies on the right of the vertical line through the Lifschitz point at $h=0, \kappa=0.5$.
Let $M$ denote the optimal observable for the order parameter discovery problem in \eqref{eq:ord-param-qcqp}, where the details concerning its solution are expanded in \myappendixref{section:ord-param-solution}. Optimizing for a single site observable, we find that $M \approx \mathds{I} - \sigma_x$, which can be interpreted as magnetization. While this observable could be used as an order parameter for the Ising-like transition, highlighted in green in \autoref{fig:phasesd-annni-i}, it failed to detect other transitions present in the ANNNI model, including the KT and PT transitions, highlighted orange and blue in \autoref{fig:phasesd-annni-i} respectively. As a result, to capture more transitions one must make use of \autoref{section:ord-param} and instead uncover a multi-site observable to detect all phase transitions present. 

For this, we expand our RDM to the two middle spin sites of the chain of length $L$. The objective is to obtain the order parameters for the paramagnetic and the anti phases, which is highlighted in \autoref{fig:phasesd-annni-i}, where we plot the result for the phase diagram construction (\autoref{section:main}) concerning the region $\mathcal{R}$, of which was obtained in the previous section using the RFS vector field.

\begin{figure}
    \centering
    \includegraphics[scale=0.23]{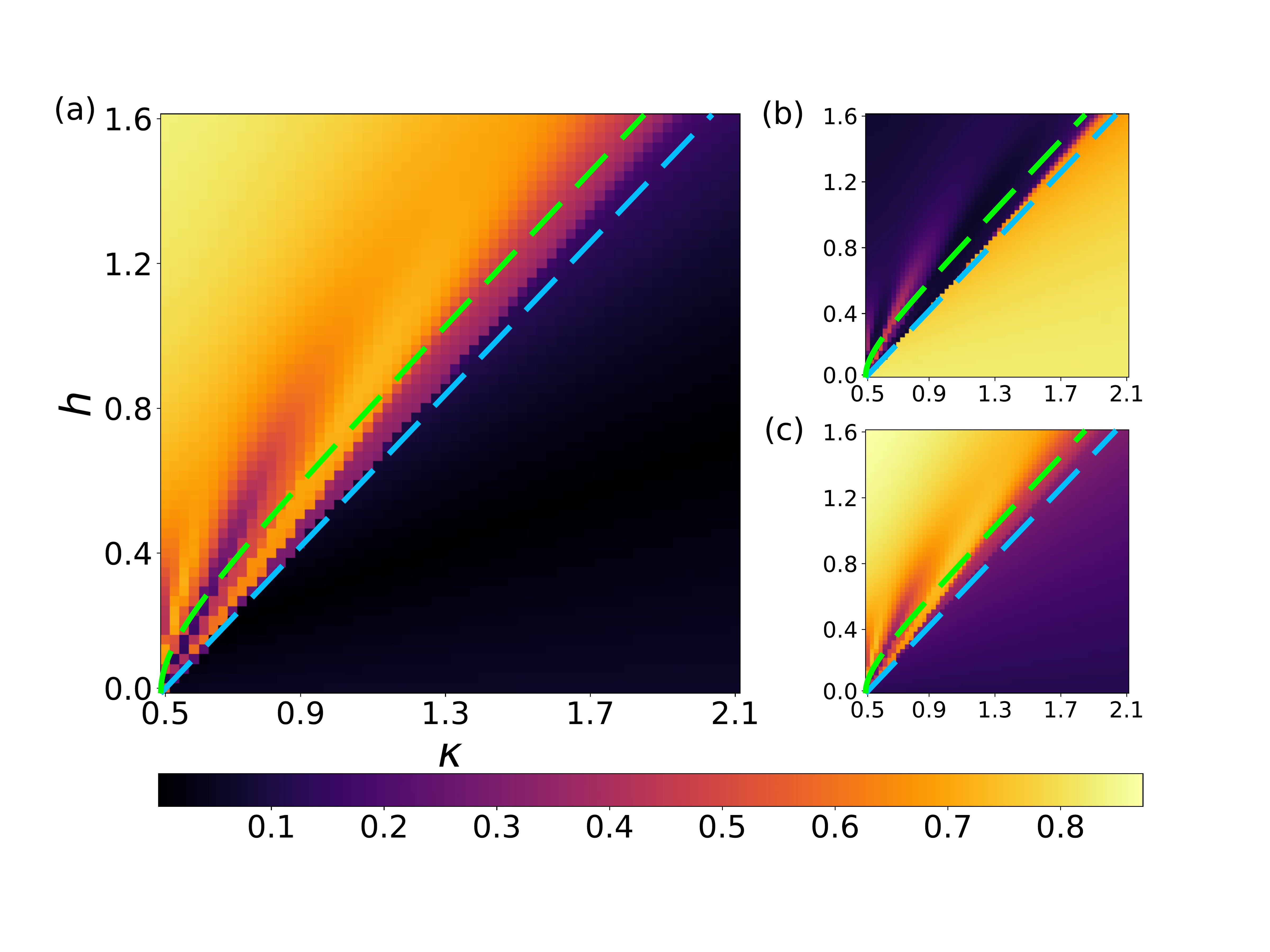}
    \caption{
    Experiments for the order parameter discovery on the ANNNI model. \newnewpart{Dashed lines indicate the known KT, and PT phase boundaries}. Here a chain length of $L = 50$ spin sites was used and a \newnewpart{two-site RDM} was used when calculating the gradient of the reduced fidelity susceptibility.
    (a) Expectation of the optimal observable $M$ for the paramagnetic phase.
    The expectations for the projectors $M^{(1)}$ and $M^{(4)}$ of \eqref{eq:annni-optim-obs-eigd}, are respectively in (c) and (b).}
    \label{fig:ordp-eigvecs}
\end{figure}

In \autoref{fig:ordp-eigvecs} (a) we present the expectations of the observable $M$ applied to the RDMs $\rho_0(\kappa_i, h_i)$,
for a finite lattice of parameters $\{(\kappa_i, h_i)\}$, following the optimization process of obtaining a relevant order parameter $M$ for this phase transition. We note that in the optimization process, the entire region, $(\kappa\quad h) \in \mathcal{R}=[0.0, 2.1] \times [0, 1.6]$, was used, which included all phase transitions present.

To understand the structure of the obtained observable, we next perform the eigendecomposition of $M$, that is
\begin{align}
    \label{eq:annni-optim-obs-eigd}
    M =& \sum_{i=1}^m \alpha_i M^{(i)}
\end{align}
where $M^{(i)}$ are rank-1 projectors and $\alpha_i$ the corresponding eigenvalues.
Let $\ket{\varphi(\theta)}=\cos(\theta)\ket{0} + \sin(\theta) \ket{1}$, we define the parametric Hermitian $B(\theta_1, \theta_2)$ as
\begin{align}
    B(\theta_1, \theta_2) \coloneqq& \ket{\varphi(\theta_1)}\bra{\varphi(\theta_1)} \otimes
    \ket{\varphi(\theta_2)}\bra{\varphi(\theta_2)}.
\end{align}
Such an operator can be interpreted as the orthogonal projector (which is defined as a square matrix $P$ such that $P^2=P=P^{\dagger})$ generated by the state 
$\ket{
    \tikz[line width=0.7pt, baseline=2pt]{\draw[->] (0, 0) -- (75:1em);
    \draw[->] (0.5em, 0) -- +(55:1em);}
}$, where the angles of the spins are $\theta_1$ and $\theta_2$, respectively.

The component corresponding to the eigenvalue with the greatest magnitude is the projector $M^{(1)}=B(\theta_1, \theta_2)$ with $\theta_1\approx 0.095 \pi$ and $\theta_2\approx 0.034 \pi$, that is $M^{(1)}\approx \ket{\uparrow\uparrow}\bra{\uparrow\uparrow}$. In \autoref{fig:ordp-eigvecs} (c), we plot the expectation of the observable $M^{(1)}$. A comparison with the plot in \autoref{fig:ordp-eigvecs} (a), shows that this is the main component for the paramagnetic phase.

The component $M^{(4)}$, with $\langle M^{(4)}\rangle$ depicted in \autoref{fig:ordp-eigvecs} (b), can be interpreted as the complementary to $M^{(1)}$. The operator approaches the following form
\begin{align}
    \label{eq:experiments-ord-param-mi}
    M^{(4)} \approx& \frac{1}{4}\left(\idenm{2}^{\otimes 2} + \paulix^{\otimes 2}\right)
    \left(\idenm{2}^{\otimes 2} - \pauliz^{\otimes 2}\right)
    = \ket{\Psi^+}\bra{\Psi^+},
\end{align}
with $\ket{\Psi^+}=(\ket{\uparrow \downarrow}+\ket{\downarrow \uparrow})/\sqrt{2}$ (Bell's state), so the ordered phase for $M^{(4)}$ is the antiphase. This can be justified by the modulated structure of the antiphase $\ket{\cdots \leftarrow \leftarrow \rightarrow \rightarrow \cdots}$ (see \myappendixref{sec:annnitheory} for a comprehensive introduction to ANNNI). Indeed we see that $\traceop\left(M^{(4)}\ket{\varphi \varphi}\bra{\varphi \varphi}\right)\ne 0$ for $\ket{\varphi}=\ket{\leftarrow}$ and $\ket{\varphi}=\ket{\rightarrow}$, whereas $\traceop\left(M^{(4)}\ket{\uparrow \uparrow}\bra{\uparrow\uparrow}\right)=0$ (paramagnetic).

\begin{figure}
    \centering
    \includegraphics[scale=0.20]{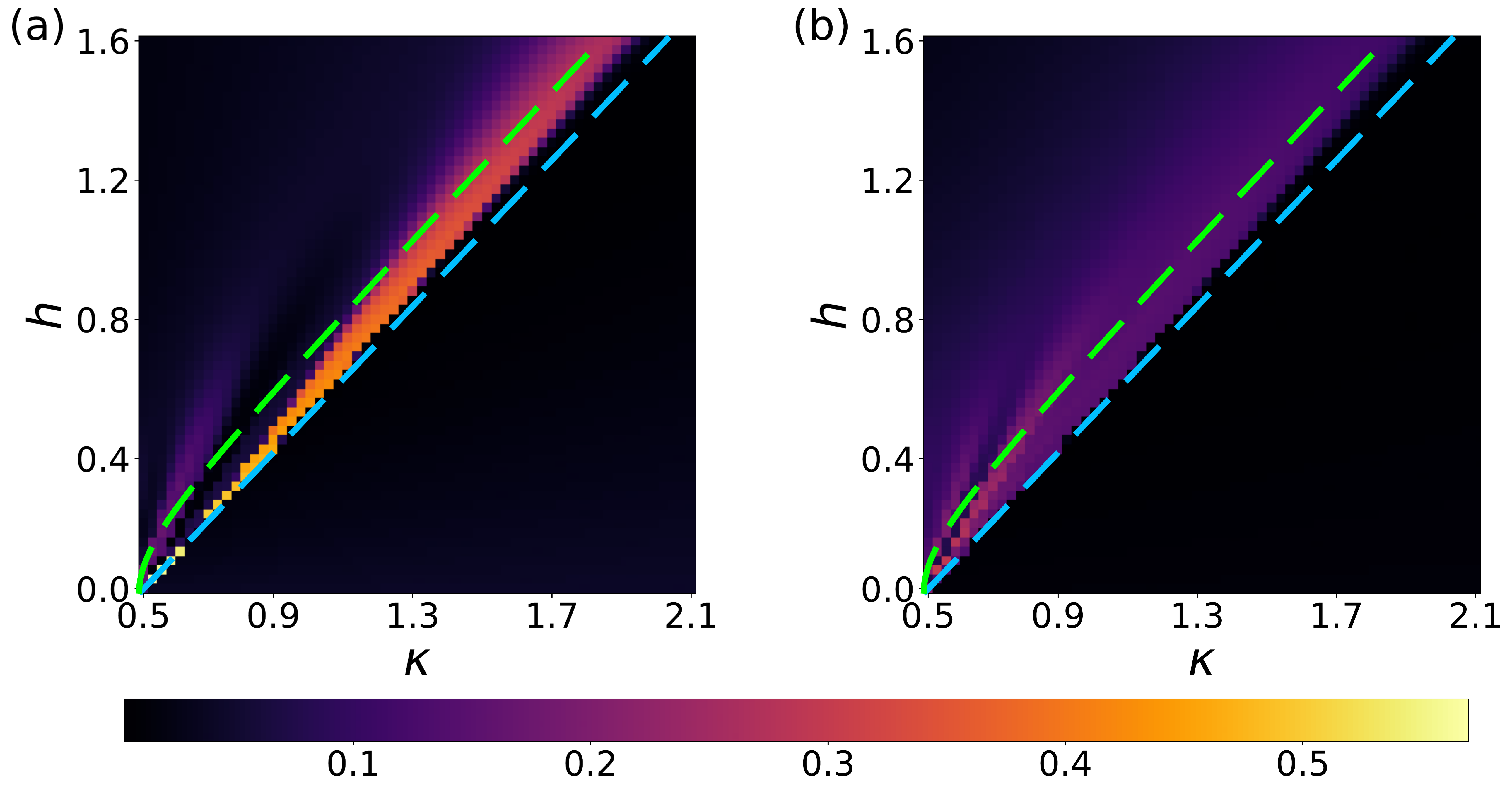}
    \caption{
    Expectations of the projectors $M^{(3)}$ (a) and $M^{(2)}$ (b) revealing the floating phase. \newnewpart{Dashed lines indicate the known KT, and PT phase boundaries }. Here a chain length of $L = 50$ spin sites was used and a \newnewpart{two-site RDM} was used when calculating the gradient of the reduced fidelity susceptibility.}
    \label{fig:ordp-eigvecs-flo}
\end{figure}

The remaining two components of the decomposition in \eqref{eq:annni-optim-obs-eigd} reveal an interesting structure. The third eigenvector determines the projector $M^{(3)}=B(\theta_1, \theta_2)$ with $\theta_1\approx 0.381 \pi$ and $\theta_2\approx -0.331 \pi$. Its expectation depicted in \autoref{fig:ordp-eigvecs-flo} (a), appears to highlight a section of the floating phase. The last projector $M^{(2)}$, whose details are omitted, produces another detail of the floating phase, which is depicted in \autoref{fig:ordp-eigvecs-flo} (b).

\subsection{Finite Size Scaling}
\label{section:fss}
Next, we validate that the obtained observable $M$ (\autoref{section:experiments-ord-params}) is indeed an order parameter for the ANNNI Model.
We accomplish this by applying finite-size scaling and verifying that the critical exponents match those expected for the Ising-like universality class. Finite-size scaling is a technique used to study phase transitions by examining how physical quantities change with system size. It involves scaling the system size and observing how properties such as critical exponents converge to their thermodynamic limits as the size increases. By applying this method, we can extrapolate results obtained from finite systems to infer behavior in the infinite system limit.
To demonstrate this phenomenon, we thus analyze the impact of chain length at the Ising-like transition. We again use the RDM $\rho_L$ for two-spin site $\{L / 2, L / 2 - 1\}$.
We calculate the expectations of the observable $M$ with the reduced density matrices $\rho_L(\kappa, h_i)$ with $\kappa = 0.001$ and $h \in [0.8, 1.2]$ \newnewpart{with coupling step $\delta = 0.001$}. In this parameter range, we can consider the next-nearest-neighbor interaction as a perturbation of the Ising model. Consequently, we anticipate observing the critical exponents characteristic of the 1D quantum Ising model.

\begin{figure}[h]
    \centering
    \includegraphics[scale=0.46]{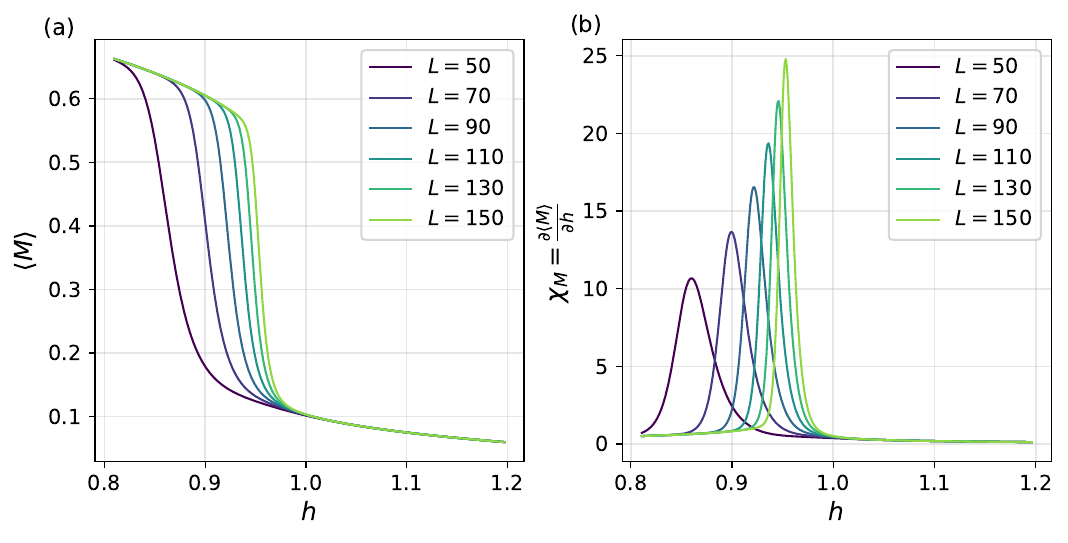}
    \caption{The \newnewpart{two-site observable} obtained using the order parameter discovery framework (a) and its gradient with $h$ (b) are applied to the ANNNI model's Ising-like transition for various chain lengths $L$. We set $\kappa = 0.001$ so that the next-nearest-neighbor term can be treated as a perturbation to the Ising model, which leads us to expect the system to belong to the Ising universality class.
}
    \label{fig:fss1}
\end{figure}

\newnewpart{From statistical mechanics (see e.g. Chapter~3 of \cite{DiFrancesco:1997nk}), we know that
\begin{align}
    \hat{M} \sim& \left|h_c^L - h_c^{\infty}\right|^{\beta} \label{eq:mag_scaling}\\
    \chi_{\hat{M}} =& \frac{d\hat{M}}{dh} \sim \left|h_c^L - h_c^{\infty}\right|^{-\gamma}
    \label{eq:sus_scaling}
\end{align}
where $\hat{M}$ is the magnetization obtained as the first derivative of the free energy $df/dh$, and consequently, the magnetic susceptibility $\chi_M$ is its second derivative $d^2 f / dh^2$. At criticality, the correlation length diverges and its scaling $\xi \sim \left| h_c^L - h_c^{\infty} \right|^{-\nu}$. A direct consequence of this law is that for a finite chain of length $L$ this relation saturates $\xi \sim L^{-\nu}$.
In \autoref{fig:fss1} we plot the expectation (with observable $M$) and its derivative, for various chain lengths $L$.
We then proceed to determine the critical transition point $h_c$ by analyzing the peak locations of the gradient of our order parameter $M$ with the assumption of behaving like \eqref{eq:sus_scaling}.
From \eqref{eq:sus_scaling} and \eqref{eq:mag_scaling} inserting the saturated correlation length just mentioned we arrive at
\begin{align}
    h_c^L =& h_c^{\infty} + a \left(\frac{1}{L}\right)^{\frac{1}{\nu}} \label{eq:crit_point_vu_crit_exp}\\
    M\left(h_c^L\right) =& a L^{-\frac{\beta}{\nu}} + b.
    \label{eq:beta_crit_exp}
\end{align}
These two relations allow us to fit the quantities shown in \autoref{fig:fss1} to perform finite-size scaling and confirming the hypothesis of the observable $M$ given by our method of \autoref{section:ord-param} to be a valid order parameter for the Ising-like transition.}

\begin{figure}[h]
    \centering
    \includegraphics[width=1.02\linewidth]{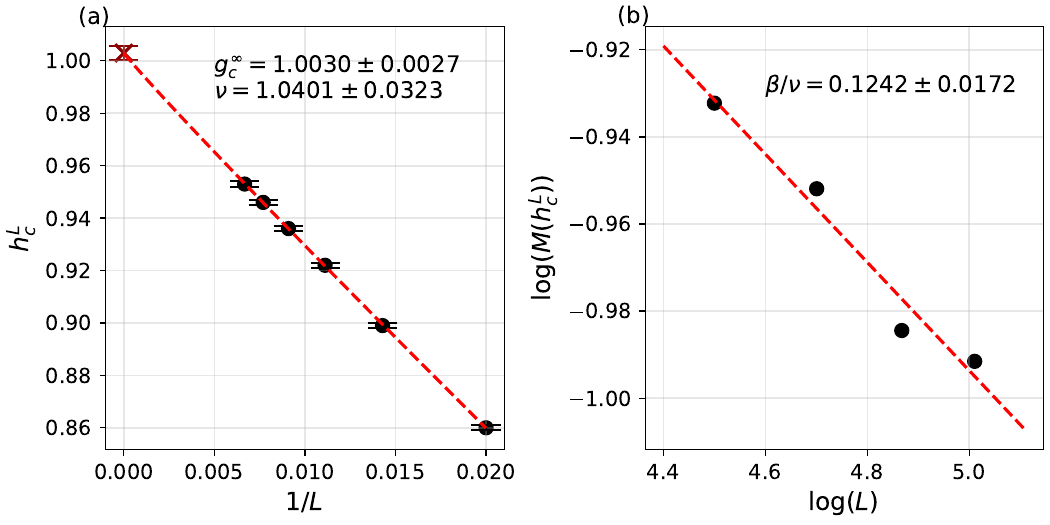}
    \caption{\newnewpart{(a) The peaks location relative to the maximum value of the gradient of the one-site observable obtained using the order parameter discovery framework is plotted against $1/L$.} A linear relationship is observed, confirming that the system remains in the Ising universality class. \newnewpart{By fitting \eqref{eq:crit_point_vu_crit_exp}, we report the critical coupling and the critical exponent relative to the correlation length. In (b), we show the log-log fit following \eqref{eq:beta_crit_exp}. The slope of the linear fit corresponds to the critical exponent ratio $\beta / \nu$.}}
    \label{fig:fss2}
\end{figure}

\newnewpart{In \autoref{fig:fss2}, we plot the fits attained from \eqref{eq:crit_point_vu_crit_exp},\eqref{eq:beta_crit_exp}.
In \autoref{fig:fss2}.a we show a power law fit which collapses to a quasi-linear one since the extracted critical value of $\nu = 1.040 \pm 0.032$ is very close to one, that is, we belong to the Ising universality class.
As we fitted the peaks location of $dM/dh$, the derivative of our observable $M$, versus the inverse of the chain length $L$, we can easily extract the peak location for the thermodynamic limit $L\rightarrow \infty$. Thus we extract a critical transition point of $h_c^{\infty}=1.003 \pm 0.003$.
Further, in \autoref{fig:fss2}.b, by applying the log function to both sides of \eqref{eq:beta_crit_exp} we show a linear fit with slope $\beta/\nu = 0.124 \pm 0.017$, perfectly in line with the theoretical value of $\beta_{\text{th}} = 1/8$ for the Ising universality class.}

\newnewpart{
Repeating the fits while varying the DMRG bond dimension, the parameter-grid spacing, and the range of system sizes included changes the exponents by less than $0.032$, which we quote as the systematic uncertainty.}

\newnewpart{By repeating the same method for a one-site rdm, the observable reproduce the same results as expected since a valid order parameter for the TFIM is the magnetization per site.} Therefore, near the Ising transition, \newnewpart{our observable} retains the critical exponents characteristic of the Ising model, \newnewpart{the latter coming from the finite-size scaling analysis}. Based on this analysis, we conclude that the obtained observable $M$ functions as an order parameter.

\subsection{The cluster Hamiltonian}
\label{section:cluster}
Following the construction of the phase diagram for the ANNNI model, we next demonstrate the versatility of the fidelity vector field method by applying it to other models. Specifically, we apply the method described in \autoref{section:main} to the cluster Hamiltonian, defined by the Hamiltonian:
\begin{align}
    H =&
    - h \sum_{i=1}^N \sigma_i^z
    - K \sum_{i=1}^{N-2} \sigma_i^x \sigma_{i+1}^z \sigma_{i+2}^x,
\end{align}
Concerning the general Hamiltonian in \eqref{eq:hamil}, the parameters $\lambda_1, \lambda_2$ are now associated with the parameters $h, K$.
When the two parameters are equal, a phase transition is expected between a trivial phase (for \( h > K \)) and a Symmetry Protected Topological (SPT) phase (for \( h < K \)). To distinguish between these two phases, string order parameters (SOPs) are commonly used, as they can reveal hidden orders or symmetries within the system. In the case of the cluster Hamiltonian, the system exhibits certain symmetries, denoted by $G_{odd} = \prod_{k=0} Z_{2 k + 1}$ and $G_{even} = \prod_{k=0} Z_{2 k}$.

From \cite{sop1, sop2}, a general SOP can be expressed as 

\begin{equation}
S_{\Sigma}^{O^L, O^R} = \lim_{|j-k| \to \infty} \left\langle \psi_0 \left| O^L(j) \left( \prod_{i=j+1}^{k-1} \Sigma_i \right) O^R(k) \right| \psi_0 \right\rangle
\end{equation}

where $\psi_0$ is the state of the system, $ \prod_{i=j+1}^{k-1} \Sigma_i$ (in our case $G_{odd}$ or $G_{even}$) preserves the local symmetries of the system, and $O^{L/R}$ are appropriately chosen to detect the phase transition. In this particular system, a typical order parameter used for the phase transition is $\sigma^x \otimes \sigma^z \otimes I \otimes \sigma^z \otimes \sigma^x$, or another commonly used one is $\sigma^x \otimes \sigma^y \otimes \sigma^z \otimes \sigma^y \otimes \sigma^z$ as discussed in \cite{sop3}.

\begin{figure}[h!]
    \centering
    \includegraphics[width=1.0\linewidth]{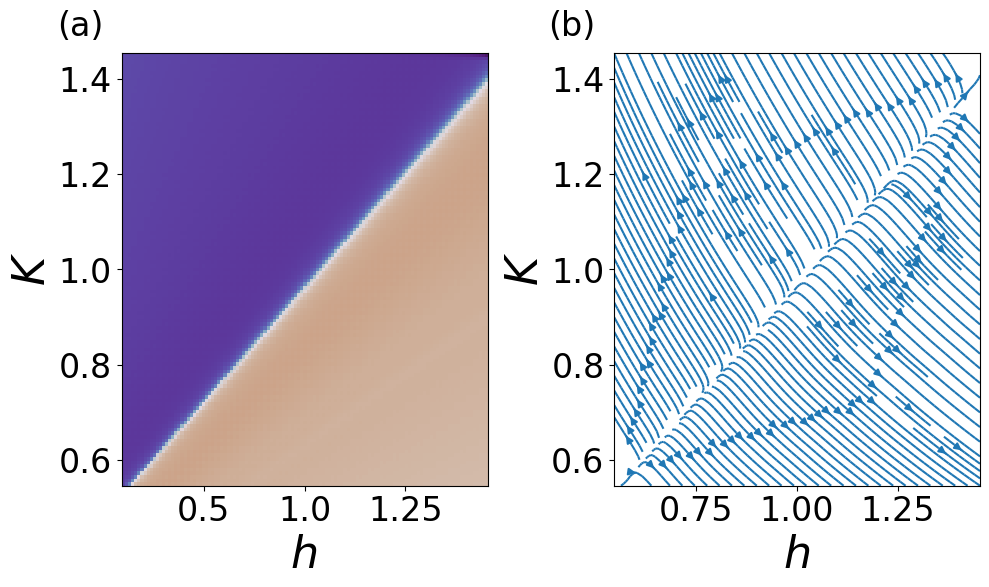}
    \caption{Phase Diagram obtained using \newnewpart{the five-site RDM when calculating the gradient of the} reduced fidelity susceptibility of the one-dimensional cluster Hamiltonian, where (a) the angle of the vector field given in \eqref{eq:angle-grad} is plotted, (b) the vector field given in \eqref{eq:vfield-def} is plotted. We note that the phase transitions correspond to a source in the vector field.}
    \label{fig:cluster}
\end{figure}

We begin by focusing on reconstructing the phase diagram as discussed in section \autoref{section:main}. To determine the ground states of the Hamiltonian, we employ DMRG, following the procedure described in \myappendixref{section:dmrg_computation}. We replicate the calculations outlined in \autoref{section:annni} to construct the phase diagram for the cluster Hamiltonian, which is presented in \autoref{fig:cluster}. The phase transition is visible, with a distinct boundary separating the two phases.

\begin{figure}[h!]
    \centering
    \includegraphics[width=1.0\linewidth]{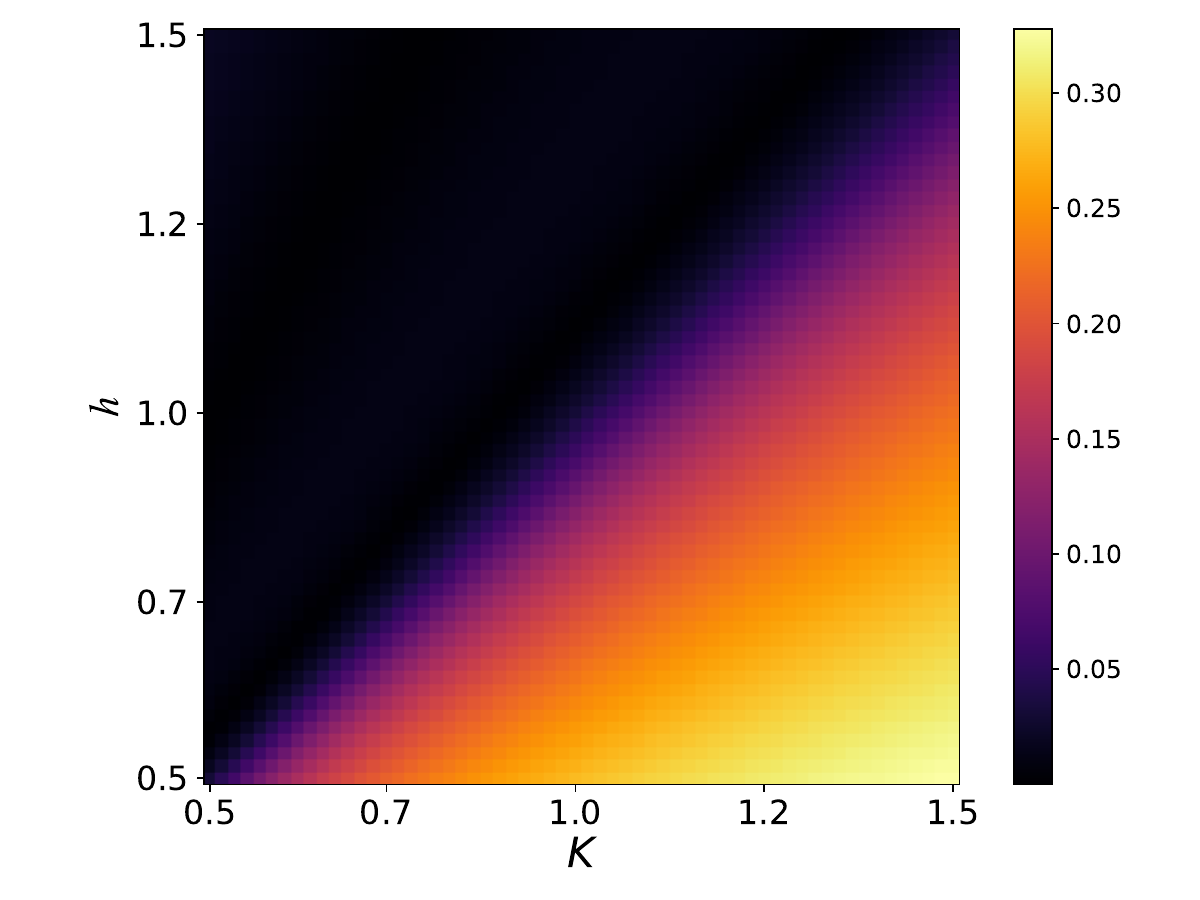}
    \caption{Expectation of the optimal observable $M$ for cluster Hamiltonian. Here, a chain length of $L = 50$ spin sites was used, and a \newnewpart{five-site RDM} was used when calculating the gradient of the reduced fidelity susceptibility.}
    \label{fig:cluster_opd}
\end{figure}

Next, we turn our attention to identifying an order parameter for this phase transition by following the method described in \autoref{section:ord-param}.
Using a five-site RDM, we derive an order parameter that effectively distinguishes between the different phases, as shown in \ref{fig:cluster_opd}. 

Although this order parameter is a linear combination of various tensor products of Pauli operators, the two terms with the largest coefficients are $\sigma^x \otimes \sigma^z \otimes I \otimes \sigma^z \otimes \sigma^x$ and $\sigma^x \otimes \sigma^y \otimes \sigma^z \otimes \sigma^y \otimes \sigma^z$, both of which reflect the symmetry of the SPT phase. This suggests that our optimizer has correctly identified the appropriate SOP and effectively exploited the system's symmetries to distinguish between the two phases.

It should be noted that similar calculations were performed using different sizes of RDMs, specifically one and two sites. However, these attempts failed to detect the phase transition, indicating that these smaller RDMs do not capture enough information about the phases on their own.

To conclude, we have demonstrated that the RFS vector field accurately reproduces the phase diagrams of both the ANNNI and cluster Hamiltonians, showcasing its robustness and reliability in capturing the phase transitions of diverse systems. Furthermore, we successfully identified an order parameter that leverages each system's symmetries to distinguish between different phases, further validating our approach.

\subsection{The Rydberg model spin chain}

\begin{figure}[h]
    \centering
    \includegraphics[width=0.9\linewidth]{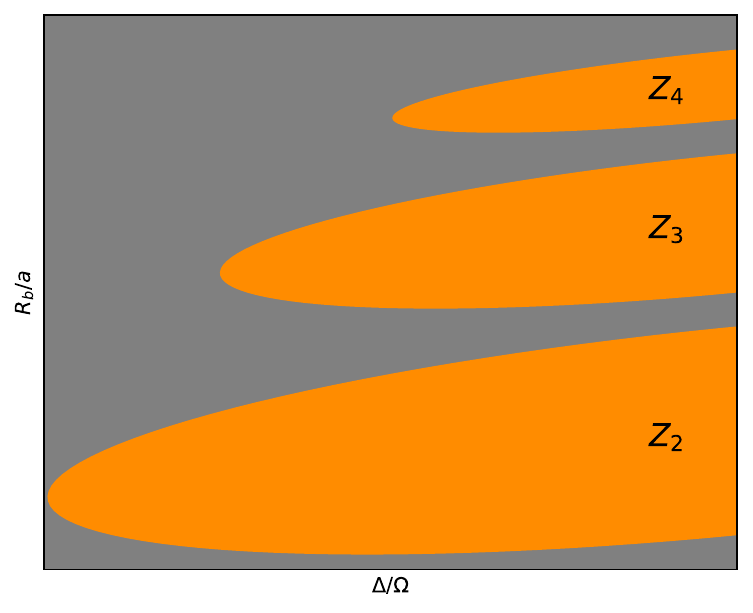}
    \caption{Schematic of the phase diagram for the Rydberg Model as a function of the parameters $\frac{\Delta}{\Omega} $ and
    $\frac{R_b}{a}$ (see the main text for more details).
    The orange regions represent the emergence of various crystalline quantum phases with $\mathbb{Z}_N$ order.}
    \label{fig:rydberg-diagram-sketch}
\end{figure}

\subsubsection{Model description}
The Rydberg model is a quantum many-body system that has attracted considerable interest, particularly in its role in experimental quantum simulation platforms
\cite{bernien2017probing, rydberg-51, rydberg-computation-i, rydberg-computation-ii}.
The system is composed of a chain of atoms such that each atom can remain in its ground state or be excited to a highly energetic Rydberg state, characterized by an electron occupying a very high orbital \cite{saffman2010quantum}. The model captures the dynamics of strongly interacting systems, as the atoms in Rydberg states exhibit long-range dipole-dipole interactions, significantly affecting the collective behavior of the system. Using these interactions, a variety of exotic quantum phases can be simulated, including quantum phase transitions and spin-liquid phases, making the Rydberg model an essential tool for probing non-equilibrium phenomena in quantum simulators \cite{browaeys2020many}.

\begin{figure*}
    \centering
    \includegraphics[width=0.8\linewidth]{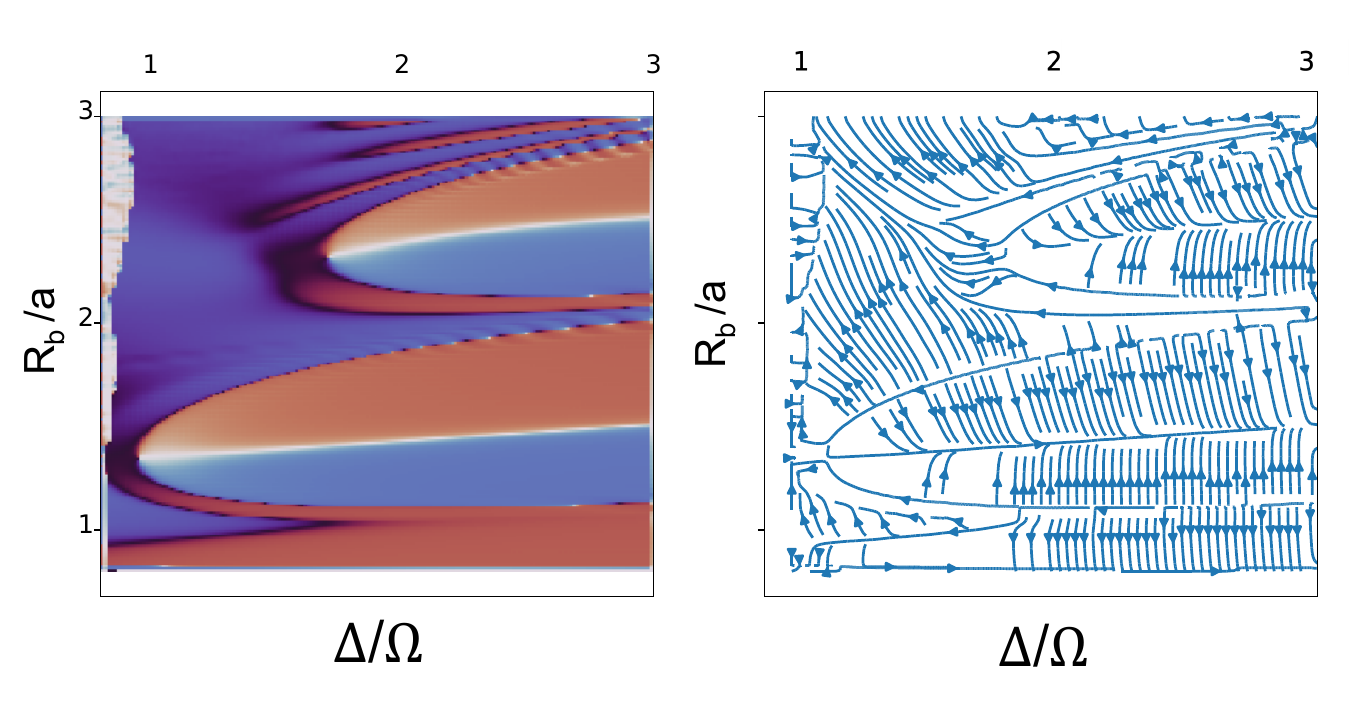}
    \caption{Phase diagram for the Rydberg Model obtained with a \newnewpart{one-site RDM} from a chain of length $L = 50$, with long-range interactions truncated at the fourth nearest neighbor. The lower crystalline phase corresponds to the $\mathbb{Z}_2$ crystalline phase, while the upper corresponds to the $\mathbb{Z}_3$ crystalline phase.}
    \label{fig:rydberg-diagram-i}
\end{figure*}

Mathematically, the Rydberg model can be formulated as a spin-$1/2$ system, where each site of the chain represents an atom that can either be in the ground state (represented by the spin state $\left| 0 \right\rangle$) or in the Rydberg state (represented by the spin state $\left| 1 \right\rangle$). The Hamiltonian of the Rydberg model is typically expressed as
\begin{equation}
    H = \Omega \sum_{i} \sigma_i^x - \Delta \sum_{i} n_i  + \sum_{i<j} V_{ij} n_i n_j,
\end{equation}
where $\sigma_i^x$ is the Pauli-x matrix acting on-site $i$, and $n_i = \left| 1 \right\rangle \left\langle 1 \right|_i$ is the number operator that projects onto the Rydberg state at site $i$. The parameters $\Omega$ and $\Delta$ represent the Rabi frequency and detuning, respectively, which control the driving field and the energy offset of the system. The second term in the Hamiltonian describes the interaction between atoms in their Rydberg states, with $V_{ij}(r) \propto 1 / r^6$ representing the strength of the interaction between atoms at sites $i$ and $j$. For \(\Delta > 0\), various spatially ordered phases emerge due to the competition between two effects: the detuning term, which favors a high population of atoms in the Rydberg state, and the Rydberg blockade, which prevents the simultaneous excitation of atoms that are closer together than the blockade radius \(R_b\), where the interaction strength \(V(R_b)\) is equal to the Rabi frequency \(\Omega\).

We define the distance between adjacent spins as $a$, thus $r = |i - j|a$, and we also define the Rydberg Radius as $V(R_b) = \Omega$. Thus, the Hamiltonian can be rewritten accordingly; 
\begin{equation}
    H =  \sum_{i} \sigma_i^x - \frac{\Delta}{\Omega} \sum_{i} n_i  + \left( \frac{R_b}{a} \right) ^6 \sum_{i<j} \frac{1}{|i - j|^6} n_i n_j,
\end{equation}
We will vary the parameters $\frac{\Delta}{\Omega} $ and $\frac{R_b}{a}$ (which correspond to $\lambda_1, \lambda_2$ in the general case in \eqref{eq:hamil})
 in the next section. 

In the limit of strong interactions, the Rydberg blockade phenomenon becomes significant, where two neighboring atoms cannot be excited to the Rydberg state due to the high energy cost. This constraint leads to the emergence of various quantum phases, including crystalline phases with $\mathbb{Z}_N$ symmetry ($N$ being an integer), which are characterized by periodic arrangements of Rydberg excitations.

A schematic of the phase diagram is given in \autoref{fig:rydberg-diagram-sketch}.

\subsubsection{Phase diagram construction}
\begin{figure}
    \centering
    \includegraphics[width=1.0\linewidth]{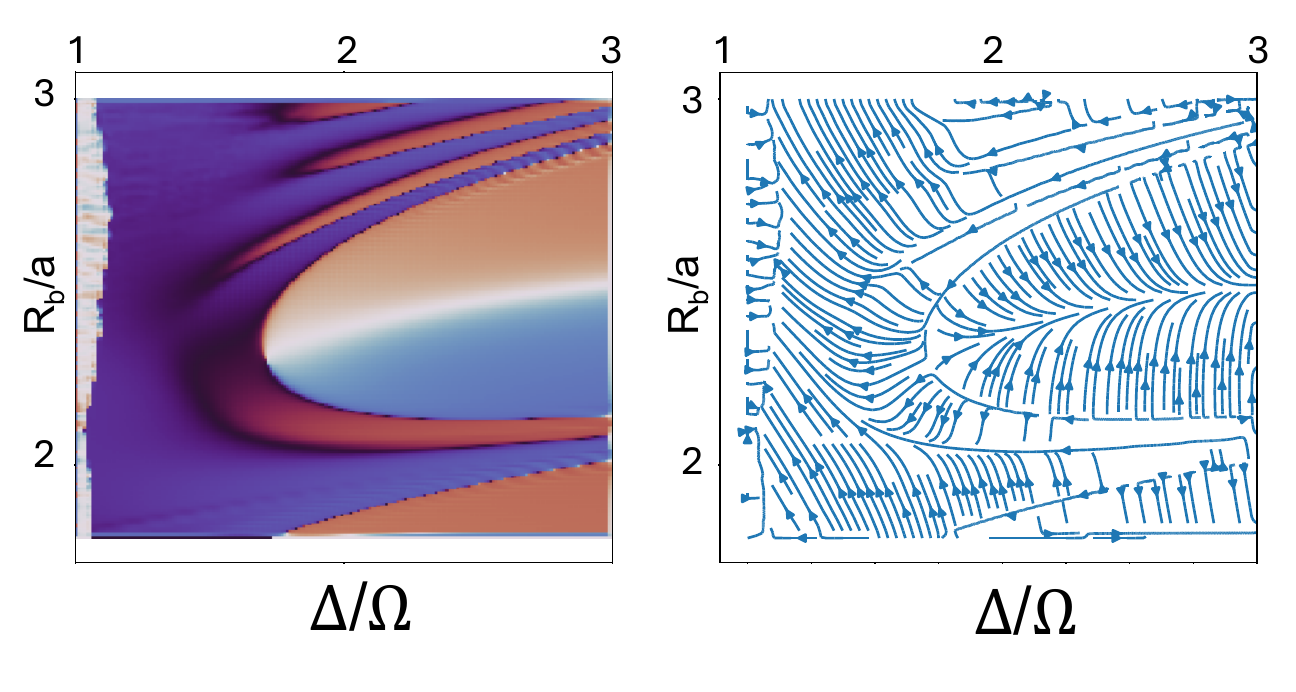}
    \caption{Detail of the phase diagram for the Rydberg Model around the $\mathbb{Z}_3$ crystalline phase \newnewpart{using a one-site RDM for the gradient of the RFS}. Due to the high resolution of the phase diagram, ripple-like features can be observed, which signal a more complex phase to be determined.} 
    \label{fig:rydberg-diagram-ii}
\end{figure}

In previous studies, the phase diagram of the Rydberg model has been explored using various techniques such as entanglement entropy \cite{PhysRevResearch.3.023049}, machine learning methods \cite{Lu:2022xof}, and fidelity \cite{PhysRevB.106.165124}. We obtain the phase diagram, with the first result presented in \autoref{fig:rydberg-diagram-i}. Here, we particularly focus on the $\mathbb{Z}_2$ and $\mathbb{Z}_3$ crystalline phases, which are typically observed in this model (See \autoref{fig:rydberg-diagram-ii} for a closer look at the $\mathbb{Z}_3$ crystalline phase). We considered a chain of length $L=50$, with long-range interactions truncated at the fourth nearest neighbor. The $1/r^6$ interaction term was approximated using an exponential form, as described in \cite{2020NatPh..16..132B}.

Our results are consistent with the parameter ranges for each crystalline phase in the phase diagram, as reported in previous studies \cite{2020NatPh..16..132B}. However, a noteworthy observation is that our method captures additional features that were not identified by earlier approaches. Specifically, we observe ripple-like features at high values of the parameters swept, which may indicate the presence of a more complex phase - a topic previously discussed in \cite{PhysRevA.98.023614, PhysRevB.106.165124, chepiga2011}. Additionally, we identify a line within each crystalline phase corresponding to the minimum of the reduced fidelity susceptibility, or equivalently, the maximum of the fidelity.

Notably, due to the high resolution of our phase diagram, we once again observe ripple-like features, consistent with previous findings in \cite{PhysRevB.106.165124}.

\newnewpart{These structures are consistent with the fidelity-susceptibility analysis of Ref.~\cite{PhysRevB.106.165124}, which characterizes the boundary of the $\mathbb{Z}_3$ lobe as a commensurate-incommensurate transition with an intermediate floating phase, the analogue of the ANNNI floating phase, of Berezinskii--Kosterlitz--Thouless type owing to an emergent U(1) symmetry \cite{2019arXiv190309179V}. The sources of the RFS vector field accumulating along this boundary thus have the same physical origin as the features reported in Ref.~\cite{PhysRevB.106.165124}.}

\subsubsection{Order parameters}
\begin{figure}
\centering
\includegraphics[width=0.9\linewidth]{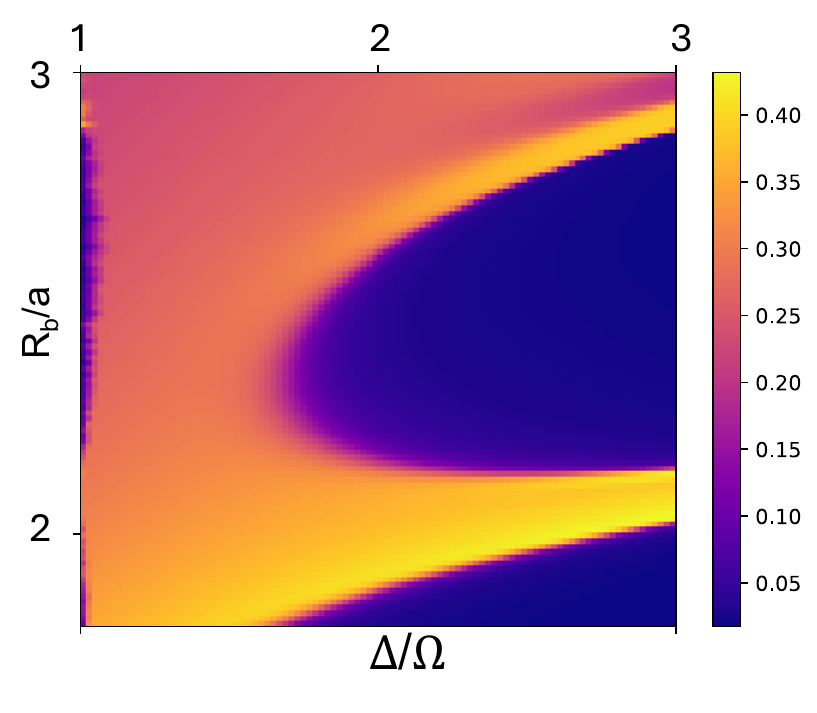}
\caption{Rydberg model:
The expectation for a \newnewpart{one-site observable} that highlights the disordered phase versus the crystalline phases $\mathbb{Z}_2$ and $\mathbb{Z}_3$.
Notably, the two crystalline phases cannot be distinguished using a \newnewpart{one-site observable}.}
\label{fig:rydberg-ordp-one-site}
\end{figure}

Next, we focus on the determination of an order parameter to detect each of these phase transitions. We initially start with a one-site RDM to try to determine the different phases as seen in \autoref{fig:rydberg-diagram-i}. The projectors from the determined order parameter can be seen in \autoref{fig:rydberg-ordp-one-site}. As is evident that while it can determine the difference between the ordered and disordered phases, it fails to differentiate the different crystalline phases. Extending this to three sites results in the same projectors, which are unable to differentiate the different crystalline phases. 

\begin{figure}[h]
\centering
\includegraphics[width=0.9\linewidth]{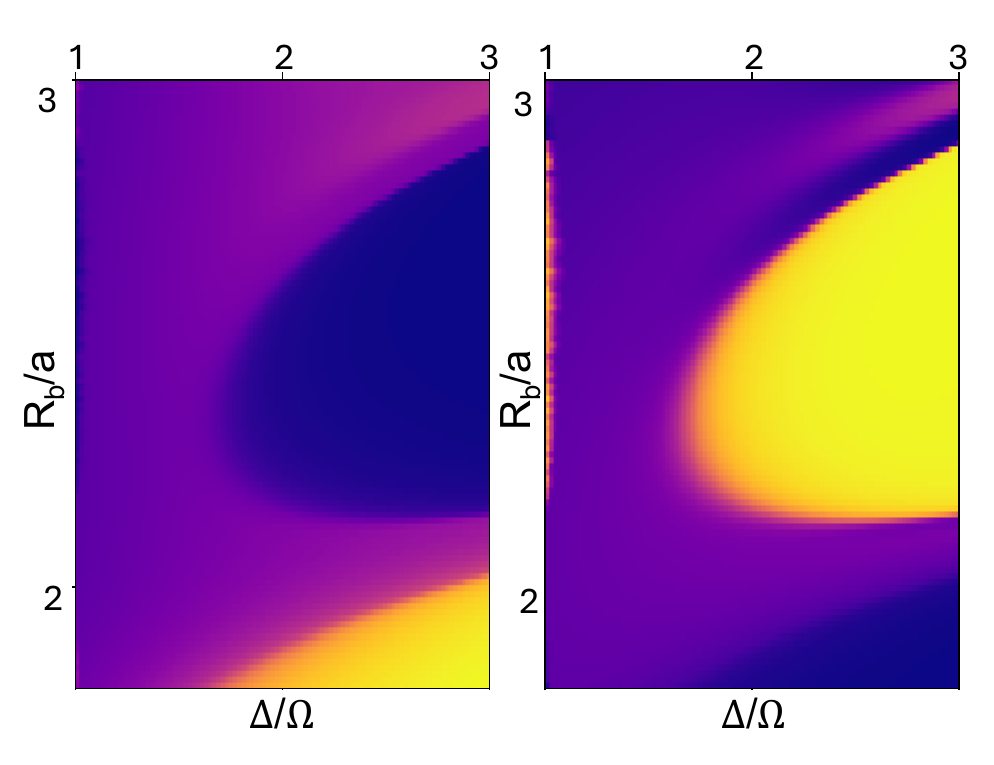}
\caption{Rydberg model: Highlights from the eigen decomposition of the order parameter on a \newnewpart{four-site RDM}.
The plots present the expectations of the projectors that reveal, respectively, the crystalline (a) $\mathbb{Z}_2$ and (b) $\mathbb{Z}_3$ phases.}
\label{fig:rydberg-ordp-four-site}
\end{figure}

It is not until a four-site RDM is used that the different disordered phases, $\mathbb{Z}_3$, $\mathbb{Z}_2$ can be characterized. This result is seen from the projectors of the observable in Figure \ref{fig:rydberg-ordp-four-site}. 

Firstly, for the $\mathbb{Z}_2$ crystalline phase, the projector of the state $\ket{0101}$ is used to detect this phase, showing that the agreed state of alternating spin up and spin down is evident in this phase, agreeing with previous literature. 

Similarly, for the $\mathbb{Z}_3$ phase, \newnewpart{the dominant projector corresponds to the state $\ket{1001}$, i.e., one Rydberg excitation every three sites within the four-site window, in agreement with the expected $\mathbb{Z}_3$ density-wave order.}
Expanding the RDM to include additional sites would make this symmetry more evident. 

\newpart{ \section{Comparison with prior methods and measurement strategies}}

\newpart{ In this section, we provide a detailed comparison between the proposed RFS vector-field framework and several existing approaches to phase detection and order-parameter identification. We focus in particular on: (i) fidelity-based methods, including subsystem fidelity approaches; (ii) operator-based discrimination methods such as Ref. \cite{PhysRevLett.96.047211}; and (iii) measurement strategies based on Pauli operators and classical shadow tomography.}

\newpart{ \subsection{Relation to fidelity and subsystem fidelity approaches}}

\newpart{ The fidelity approach to quantum phase transitions is based on the overlap between nearby ground states, $F(\lambda,\lambda+\delta) = |\langle \psi(\lambda) | \psi(\lambda+\delta) \rangle|,$ and its associated fidelity susceptibility ($\chi_F(\lambda)$), which captures the leading quadratic response. These quantities are known to exhibit universal scaling behavior near critical points and provide a basis-independent probe of phase transitions.}

\newpart{Global metrics identify critical points via non-analyticities, but they are difficult to compute and to measure for large many-body systems; moreover, global overlaps concentrate exponentially with system size \cite{thanasilp2024exponential}, and they can be ill-behaved across level crossings \cite{Kwok2008partial}. Fidelity ideas were extended beyond pure global states along two distinct lines. Mixed-state fidelity and the associated Bures metric were studied for \emph{thermal-state} manifolds by Zanardi and collaborators \cite{PhysRevA.75.032109, PhysRevA.76.062318}, primarily to generalize fidelity criticality to finite temperature. Independently, the \emph{reduced} fidelity of ground-state RDMs was developed as a probe of QPTs in its own right \cite{Zhou2007rgfidelity, Kwok2008partial, Ma2008rfsLMG, Ma2008rfsIsing, Xiong2009rfsHeisenberg}; Section VI.B of the review \cite{GU_2010} gives a comprehensive account. Partial trace preserves criticality signatures through the monotonicity of the Uhlmann fidelity under CPTP maps \cite{Mendon_a_2008}, and the reduced quantities remain accessible when global overlaps are not.}

\newnewpart{Our construction builds directly on this framework. Two remarks delimit what is, and is not, new:
\begin{itemize}
\item Metric contraction: multi-parameter fidelity susceptibilities are standard through the quantum geometric tensor \cite{VenutiZanardi2007} and the multivariate Fisher information metric (see, e.g., Eq.~(2) of Ref.~\cite{Kasatkin2024fim}). Our scalar $g(\blambda) = -\sum_k \partial_{\delta_k}^2 f(\blambda,\bdelta)\big|_{\bdelta=0}$ is the trace of the Bures metric tensor of the \emph{RDM manifold}, a specific, basis-independent contraction of this known object (\myappendixref{section:info-geo-link}), chosen because it aggregates the susceptibilities of all parameter directions into a single non-negative field suitable for the gradient construction below.
\item Vector-field structure: two-dimensional maps of the fidelity susceptibility across a phase diagram have been produced before, notably by the ClassiFIM approach \cite{Kasatkin2024fim, Kasatkin2024classifim}, which estimates the Fisher information metric over a 2D parameter space from single-basis measurements. Our contribution is the construction built on top of such a map: the gradient field $P(\blambda) = -\nabla g(\blambda)$, whose sources localize the phase transitions and whose sinks identify representative states of each phase, the scale invariance of the angle field $\theta(\blambda)$ near criticality (\myappendixref{section:scale-invariant}), and the use of the angular labels as supervision for the order-parameter optimization of Eq.~\eqref{eq:ord-param-qcqp}. We note that plotting the scalar $\traceop(h_{\blambda}) = -g(\blambda)$ (e.g. see left plot in Figs.~\ref{fig:phasesd-annni-i}-\ref{fig:cluster}-\ref{fig:rydberg-diagram-ii}) alone detects the main phases, but nearby transitions merge into a broad ridge; the vector field resolves them as distinct sources, which is what permits the identification of the floating-phase, SPT, and crystalline phases in \autoref{section:experiments}.
\end{itemize}}
In the ANNNI model, we find that:
\begin{itemize}
\item Global fidelity susceptibility ($\chi_F(\lambda)$) and subsystem fidelity both identify the primary Ising transition.
\item The RFS-derived quantity ($g(\lambda)$) exhibits similar enhancements but also resolves competing structures in multi-phase regions.
\item The vector field ($P(\lambda)$) organizes these features into a coherent flow pattern, enabling simultaneous identification of multiple transition lines without specifying perturbation directions.
\end{itemize}

\newnewpart{\emph{Scaling of the reduced susceptibility.} For the global fidelity susceptibility, the scaling $L^{-d}\chi_F \sim L^{2/\nu - d}$ holds near criticality \cite{PhysRevB.81.064418}. The reduced quantity generally obeys different, subsystem-dependent laws: the two-site RFS of the transverse-field Ising chain diverges only as $(\ln N)^2$, so that its square root is the quantity obeying standard finite-size scaling with $\nu=1$ \cite{Ma2008rfsIsing}; in the dimerized Heisenberg chain the two-site RFS follows the square of the second derivative of the ground-state energy \cite{Xiong2009rfsHeisenberg}; and in the LMG model its maximum scales as $N^{2/3}$ \cite{Ma2008rfsLMG}. We therefore do not assume the global scaling relation for $g(\blambda)$. Our working hypothesis, supported by the numerics of \autoref{section:experiments}, is that a $k$-site RDM with $k$ comparable to the range of the local terms of the driving Hamiltonian captures the critical behavior of the corresponding transition: this is verified for the Ising transition of the ANNNI model in \autoref{section:fss} ($\nu=1$, $\beta \approx 1/8$), and it motivates the choice $k=4$ for the Rydberg chain, whose truncated van der Waals interaction couples up to fourth neighbors.}

\newnewpart{\emph{When is the reduced construction advantageous?} The method requires only the $k$-site RDMs on a lattice of parameter points; the DMRG ground states used here are a data source, not an assumption. Any procedure yielding local mixed states---local tomography on a quantum simulator, classical shadows restricted to a $k$-site cluster \cite{huang2020predicting}, or simulations that never produce a global pure state---provides valid input. On experimental platforms only local measurements are typically feasible, and the exponential concentration of global overlaps \cite{thanasilp2024exponential} makes the global fidelity impractical to estimate at scale, whereas the $k$-site RDM requires at most $4^k-1$ Pauli expectation values independently of the total system size. Empirically, \autoref{fig:comparison-gfs-rfs} in \myappendixref{section:dmrg_computation} compares global and reduced susceptibilities on all three models: the two agree on the main transitions, the global quantity exhibits an artifact for the cluster model that is absent in the RFS, and the GFS resolves the Rydberg $\mathbb{Z}_2$ lobe more sharply. This is expected by construction. As mentioned in \autoref{section:practical-phase-diag-construction}, we know that $f(\boldsymbol{\lambda}, \boldsymbol{\delta}) \ge \left|
\bra{\gstate{\blambda}}
\ket{\gstate{\blambda + \bdelta}}
\right|$,
so the reduced fidelity is an upper bound for the global fidelity. It follows that the variations of the RF are upper bounded by the variations of the GF, that is
$g(\boldsymbol{\lambda}) \leq \mathcal{X}_F$ (with $\mathcal{X}_F$ defined in \eqref{eq:fidelity-sus-def}).
}

\newpart{ \subsection{Comparison with operator-based approaches (Ref. \cite{PhysRevLett.96.047211})}}

\newpart{Operator-based methods, such as Ref. \cite{PhysRevLett.96.047211}, construct observables that optimally distinguish between two reduced density matrices ($\rho_A^{(1)}$) and ($\rho_A^{(2)}$), typically by maximizing a suitable distance or distinguishability measure.}

\newpart{The construction of Furukawa, Misguich, and Oshikawa \cite{PhysRevLett.96.047211} is the direct antecedent of ours: given two representative RDMs, it finds the observable that optimally distinguishes them, and our closed-form two-state solution, Eq.~\eqref{eq:ord-param-special-em}, is precisely of this type; for the Ising transition both approaches recover an operator closely related to the conventional order parameter. A grid of such pairwise comparisons between neighboring parameter points could, in principle, also be used to chart a phase diagram---indeed our fidelity-perturbation lattice (\autoref{fig:gstates-lattice}) has exactly this local pairwise structure. The differences are threefold. (i) Aggregation: Ref.~\cite{PhysRevLett.96.047211} yields one discriminating observable per chosen pair, whereas Eq.~\eqref{eq:ord-param-qcqp} is a single joint optimization over all pairs $(i,j) \in I^+ \times I^-$ induced by the RFS labels, producing one observable whose eigendecomposition simultaneously encodes several phases (\autoref{section:experiments-ord-params}). (ii) Label origin: the labels supervising Eq.~\eqref{eq:ord-param-qcqp} are generated from the same RFS data via the angular construction of \autoref{section:ord-param}, so no external identification of phases or representative states is required. (iii) Diagnostics: the indefiniteness condition on the matrix $A$ (\myappendixref{section:ord-param-solution}) signals when the chosen subsystem size cannot support a discriminating observable. For KT and PT transitions, where phases are not distinguished by simple symmetry-breaking patterns, single pairwise comparisons are less informative; the RFS framework aggregates many local comparisons into the geometric structure of parameter space, which is what resolves these boundaries in practice (\autoref{section:annni}).}

\newpart{Overall, the RFS approach can be viewed as a generalization of pairwise operator discrimination to a global, geometry-driven setting.}

\newpart{\subsection{Comparison with Pauli-based measurements and classical shadows}}

\newpart{An important practical question concerns how the required quantities can be measured experimentally.}

\newpart{Classical shadow tomography, introduced by \cite{huang2020predicting}, as well as related operator-structure measurements discussed in \cite{8qt9-72ts}, predicts many properties of quantum systems from a few measurements and provides an efficient framework for estimating expectation values of a large number of observables. In particular, expectation values of all $k$-local Pauli operators can be estimated with several measurements scaling as $\mathcal{O}\left( \frac{k\log(n) 4^k}{\varepsilon^2} \right),$ where $n$ is the system size and $\varepsilon$ the target precision. In contrast, the RFS approach requires access to the complete characterization of the reduced density matrix of $k$-sites.}

\newnewpart{Experimentally, a $k$-site RDM can be reconstructed from its Pauli expansion, $\rho_A = 2^{-k}\sum_{P \in \{I,X,Y,Z\}^{\otimes k}} \langle P \rangle\, P$, i.e., from up to $4^k - 1$ non-trivial Pauli expectation values; for the values $k \le 5$ used here this cost is independent of the total system size, although exponential in $k$. We stress that the numerical results in this work use RDMs obtained exactly from DMRG ground states rather than finite-shot measurement data, so experimental applicability is currently a compatibility statement rather than a demonstration; a systematic finite-shot study is left for future work.}

\newnewpart{The relative measurement cost of the two approaches depends on the task and on the structure of the observables required to distinguish the phases. The RFS framework requires reconstruction of the complete $k$-site reduced density matrix, corresponding to the estimation of up to \(4^k-1\) non-trivial Pauli expectation values. We therefore do not claim that the RFS framework is generally more measurement-efficient than classical-shadow approaches. When the phases can be distinguished using a sparse set of low-bodyness Pauli observables, shadow-based methods may require substantially fewer measurements.
The distinction instead lies in the information produced by the two workflows. The RFS framework uses the reconstructed local states to construct a geometric phase diagram and subsequently optimizes over the complete Hermitian operator space of the selected subsystem, yielding an explicit observable. Classical shadows provide an efficient measurement strategy for simultaneously estimating many candidate observables and could also be used to estimate the local Pauli expectation values required to reconstruct the RDMs used by our method. The two approaches should therefore be regarded as complementary rather than competing.}

\newnewpart{Two shadow-based approaches deserve explicit comparison. First, Ref.~\cite{8qt9-72ts} performs supervised phase classification using Pauli shadows restricted to the low-bodyness operator subspace: the informative low-weight Pauli operators are identified from the labelled training data themselves, and the resulting classifier defines a dataset-dependent discriminating observable. When such a sparse set of low-bodyness observables suffices, this is substantially cheaper in measurements than reconstructing a complete $k$-site RDM, and we make no claim of a general measurement-efficiency advantage over it. The distinction lies in the workflow and the optimization domain: in our approach the phase labels are constructed without supervision from the RFS vector field, and Eq.~\eqref{eq:ord-param-qcqp} then optimizes over the complete Hermitian operator space of the $k$-site cluster---a trust-region program whose global optimum is obtained in time polynomial in the RDM dimension $2^k$, with no heuristic operator selection and no training (\myappendixref{section:ord-param-solution})---and whose two-state case admits the closed form of Eq.~\eqref{eq:ord-param-special-em}, an explicit separating observable whenever the two reduced states differ. This constitutes an advantage in optimization guarantees and interpretability, not in measurement complexity. Second, Huang \emph{et al.}~\cite{Huang2022science} use classical shadows to build a similarity (shadow) kernel and apply kernel PCA---unsupervised---or supervised kernel methods to classify quantum phases, with rigorous efficiency guarantees for broad classes of gapped phases. Their pipeline is measurement-efficient and provably efficient, and returns a classifier in an implicit kernel feature space; ours requires the complete $k$-site RDM but returns an explicit Hermitian observable together with a geometric phase diagram, with no kernel choice or training. The approaches are complementary: classical shadows can supply the RDM estimates our method consumes, and a quantitative benchmark against the shadow-kernel pipeline of Ref.~\cite{Huang2022science} is a natural next step, which we leave to future work.}

\newpart{\subsection{Summary of comparative advantages}}

\newpart{The comparisons above highlight the distinctive features of the RFS vector-field framework: (i) It retains the universality and sensitivity of fidelity-based approaches; (ii) It extends subsystem fidelity methods by incorporating directional and geometric information; (iii) It generalizes operator-based discrimination to a global, data-driven setting; (iv) It is compatible with modern measurement strategies, including classical shadows and local measurement protocols. These properties make it a versatile tool for both theoretical and experimental investigations of quantum phase structure.}

\vspace{1cm}

\section{Conclusion}

We have developed a novel and broadly applicable method for detecting quantum phase transitions based on reduced fidelity susceptibility (RFS), requiring only local reduced density matrices and no prior knowledge of symmetries, order parameters, or the nature of the transition. Applied to the axial next-nearest-neighbor Ising (ANNNI) model, our approach reveals subtle features of the phase diagram that elude conventional techniques, demonstrating both its sensitivity and efficiency. The RFS framework constructs a vector field that offers intuitive visual and conceptual access to phase structure, making it a powerful diagnostic for complex many-body phenomena. Looking forward, we aim to extend this method to systems exhibiting topological order—where traditional local order parameters fail—by exploiting its compatibility with non-linear functions of local reduced states (e.g., entanglement entropy) \cite{Hamma_2005, Levin_2006, Kitaev_2006}, thereby probing global quantum correlations through minimal local information.

Additionally, our framework demonstrates significant promise for application in quantum hardware environments. Utilizing the reduced density matrix thermodynamic information, it circumvents the need for full-function tomography while still capturing essential characteristics of phase transitions.

Moreover, we extended our framework to devise a method for discovering order parameters in systems where they are unknown. The reversed method, applied to the ANNNI model, was validated through the decomposition and finite-size scaling of the identified order parameter. This validation underscores the framework's potential to explore and analyze novel systems, including unexplored phenomena such as floating phases in a quantum system.
We believe that the proposed approach is an important milestone in phase transition detection and order parameter discovery, offering a robust tool for exploring the complex landscape of quantum systems.

\label{section:conclusion}

\section{Acknowledgment}

We are grateful to Dmytro Mishagli (IBM) and Victor Valls (IBM) for their valuable input and suggestions.

F.DM and E.R. acknowledge support from the BasQ strategy of the Department of Science, Universities, and Innovation of the Basque Government.

E.R. is supported by the grant PID2021-126273NB-I00 funded by MCIN/AEI/ 10.13039/501100011033 and by ``ERDF A way of making Europe" and the Basque Government through Grant No. IT1470-22. This work was supported by the EU via QuantERA project T-NiSQ grant PCI2022-132984 funded by MCIN/AEI/10.13039/501100011033 and by the European Union ``NextGenerationEU''/PRTR. This work has been financially supported by the Ministry of Economic Affairs and Digital Transformation of the Spanish Government through the QUANTUM ENIA project called – Quantum Spain project, and by the European Union through the Recovery, Transformation, and Resilience Plan – NextGenerationEU within the framework of the Digital Spain 2026 Agenda.

This work has been partially funded by the Eric \& Wendy Schmidt Fund for Strategic Innovation through the CERN Next Generation Triggers project under grant agreement number SIF-2023-004.

\section{Author Contributions}

All authors contributed to the conceptualization and methodology of the study and to the writing and revision of the manuscript. Large Language Model (AI) tools were used for linguistic style refinement and to assist in developing part of the code used to reproduce plots and to make the released software package more user-friendly.

\newnewpart{\section{Code and Data Availability}}

\newnewpart{All the simulations and data generated for this manuscript are available at the github \cite{qupytex-repo} repository and Zenodo \cite{zenodo-repo} repository, respectively.}

\vspace{1cm}

\appendix
\section{Construction of the phase diagram}
\label{section:practical-phase-diag-construction}
\begin{figure}
    \centering
    \begin{tikzpicture}[scale=1.5]
    \draw[draw=red, dashed] (0,0) rectangle ++(3,2);
    \draw[->, thick] (0, 0) -- (3.5, 0) node[right] {$\lambda_1$};
    \draw[->, thick] (0, 0) -- (0, 2.5) node[above] {$\lambda_2$};

    \foreach \x in {0, 1, 2, 3} {
    \foreach \y in {0, 1, 2} {
      \fill (\x, \y) circle (2pt);
    }
    }
    
    \foreach \x in {0.5, 1.5, 2.5} {
    \foreach \y in {0, 1, 2} {
      \fill[red] (\x, \y) circle (1pt);
      \draw[red] (\x, \y) circle (2pt);
    }
    }
    
    \foreach \x in {0, 1, 2, 3} {
    \foreach \y in {0.5, 1.5} {
      \fill[red] (\x, \y) circle (1pt);
      \draw[red] (\x, \y) circle (2pt);
    }
    }

    \node[right] at (3, 1) {$\scriptstyle \rho_0^{i, j + 1}$};
    \node[right] at (3, 2) {$\scriptstyle \rho_0^{i+1, j + 1}$};

    \node (rho1) at (2, 2.5) {$\scriptstyle \rho_0^{i, j}$};
    \node (d1) at (2.5, -0.5) {$\scriptstyle f(\boldsymbol{\lambda}_{i, j}, h\mathbf{e}_1)$};
    \node (d2) at (0.5, -0.5) {$\scriptstyle f(\boldsymbol{\lambda}_{i, j}, -h\mathbf{e}_1)$};
    \draw[->, densely dashed, bend left=30] (2, 1) to (rho1);
    \draw[->, densely dashed, bend left=30] (2.5, 1) to (d1);
    \draw[->, densely dashed, bend left=30] (1.5, 1) to (d2);
\end{tikzpicture}
    \caption{Structure of the lattice of RDMs $\rho_0^{i, j}\coloneqq\rho_0(\boldsymbol{\lambda}_{i, j})$ induced by a finite lattice
    of parameters $\{ \boldsymbol{\lambda}_{i, j} \}$, contained in a region $\mathcal{R}$ (outer rectangle).}
    \label{fig:gstates-lattice}
\end{figure}

Starting from results devised in \autoref{section:main}, we proceed with the numerical construction of the phase diagram.
We consider a rectangular region of the Hamiltonian parameters $\mathcal{R} \subseteq \mathcal{X}$.
Given a step in the parameters space $h>0$, we construct a finite lattice of parameters $\{\blambda_{i, j}\} \subset \mathcal{R}$, such that
\begin{align}
    \blambda_{i+d_i, j+d_j}=\blambda_{i, j} + d_j h \bvecnodim{1} + d_i h \bvecnodim{2},
\end{align}
with $d_i, d_j$ belonging to a subset of $\mathbb{Z}$.
Using the density matrix renormalization group (DMRG) \cite{RevModPhys.77.259, tenpy},
we obtain the RDM related to the ground state (for a selected subsystem) for the parameter $\blambda_{i, j}$, which we define as
$\rho_0^{i, j}=\rho_0(\blambda_{i, j})$, where the RHS is defined in \eqref{eq:gstate-rho}.
In \autoref{fig:gstates-lattice} the lattice points for the RDMs $\rho_0^{i, j}$ are represented by the symbol \tikz{\fill (0, 0) circle (2pt);}.

We proceed with the preparation of an approximation $\tilde{g}$ of the function $g$, that is the RFS defined in \eqref{eq:fun-g-def}.
The latter is the Laplacian of the function $f(\blambda, \bdelta)$ defined in \eqref{eq:fun-f-def}.
So, starting from the lattice of parameters $\{\blambda_{i, j}\}$, we obtain the fidelity perturbations
\begin{align}
    f(\blambda_{i, j}, d_j h \bvecnodim{1} + d_i h \bvecnodim{2}) = \sqrt{F}\left(\rho_0^{i, j}, \rho_0^{i + d_i, j + d_j}\right),
\end{align}
for a fixed step $h$ (which defines the lattice of parameters) and $d_i, d_j \in \{-1, 0, 1\}$.
We can immediately verify the computational advantage of the approach since adjacent
lattice points share a fidelity perturbation.
For example $f(\blambda_{i, j}, h \bvecnodim{1})=f(\blambda_{i, j+1}, - h \bvecnodim{1})$.
In \autoref{fig:gstates-lattice}, the fidelity perturbations are represented with the symbol \tikz{
    \fill[red] (0, 0) circle (1pt);
    \draw[red] (0, 0) circle (2pt);
} and placed between adjacent lattice points to emphasize the concept of sharing.
By considering the finite differences approximation of the second derivative\footnote{
    For a function $f:\mathbb{R} \to \mathbb{R}$, under sufficient smoothing conditions, we make use of the following approximation
    \begin{align}
        \frac{\mathrm{d}^2f}{\mathrm{d}x^2}\approx& \frac{f(x+h) + f(x-h) -2f(x)}{h^2},
        \label{eq:second-discr-der}
    \end{align}
    for some $h>0$.
}, and by noting that $f(\blambda_{i, j}, \mathbf{0})=1$ (for all valid $i, j$), we obtain
\begin{align}
    \tilde{g}(\blambda_{i, j}) =& \frac{4 - \sum_{\bdelta \in \Omega}f(\blambda_{i, j}, \bdelta)}{h^2},
\end{align}
where $\Omega=\{\pm h\bvecnodim{1}, \pm h\bvecnodim{2}\}$ is the set of perturbation displacements around $\blambda_{i, j}$.
In practice, we omit the factor $1/h^2$ (numerically convenient) which is irrelevant for the subsequent computations.
We proceed by adhering to the structure outlined in \autoref{section:main}. Obtained the RFS approximation $\tilde{g}$, we continue with the computation of its gradient. Before that, we need to introduce a few concepts related to discrete signal processing.

For a continuous function $f:\mathbb{R}^2 \to \mathbb{R}^2$, we define the convolution at $(x, y) \in \mathbb{R}^2$ for discrete 2-dimensional signals as
\begin{align}
    f(x, y) * k \coloneqq \sum_{i=-N}^N\sum_{j=-N}^N k(j, i) f(x-j h, y-i h),
\end{align}
where $k$ is a convolution kernel with support $N$, that is $k(j, i)=0$ for $|i|>N$ or $|j|>N$. Also, the scalar $h>0$ represents the step for the lattice of points $\{(x\ y)^{\top} + (jh\ ih)^{\top}|i, j \in \mathbb{Z}\}$.
We introduce a discrete differentiation operator called the \textit{Sobel operator} \cite{sobel} whose $x$ component is given by the kernel matrix
\begin{align}
    G_x \coloneqq \begin{pmatrix}
    -1 & 0 & 1\\
    -2 & 0 & 2\\
    -1 & 0 & 1
    \end{pmatrix}.
\end{align}
The Sobel finds applications mainly in computer vision and it was initially developed to obtain an efficiently computable gradient operator with more isotropic characteristics than the Roberts cross operator \cite{edge-det-survey}.

Now, we define our kernel as $k=G_x + \imath G_x^{\top}$, which corresponds to the approximation of the gradient w.r.t. $x$ on the real part and the $y$ component on the imaginary part.

We obtain an equivalent approximation of $P$ in \eqref{eq:vfield-def} using the convolution
\begin{subequations}
\begin{align}
    \tilde{P}(\blambda_{i, j}) =& -\tilde{g}(\blambda_{i, j}) * k\\
    =& -\sum_{a=-1}^1\sum_{b=-1}^1 k(a, b) \tilde{g}(\blambda_{i - a, j - b}) \in \mathbb{C}.
\end{align}
\end{subequations}
When the indices $(i, j)$ of the lattice elements are beyond the limits of definition we define $\tilde{g}(\blambda_{i,j})=0$.

The first outcome of the process is a point-to-color graph
\begin{align}
    \blambda_{i, j} \mapsto c\left(\mathrm{Arg}\left(\tilde{P}(\blambda_{i, j})\right)\right),
\end{align}
where $c(\cdot)$ is a colormap which we introduce now. Let $\mathcal{C}$ be a space of colors and let $c: (-\pi, \pi] \to \mathcal{C}$ be a mapping from the angle $\theta$ to a color in $\mathcal{C}$.
The function $c$ is required to be smooth on $(-\pi, \pi)$ and non-constant, also it must be such that $\lim_{\theta \to -\pi}c(\theta)=c(\pi)$. The latter point is fundamental for dealing with the discontinuity of $\mathrm{Arg}(\cdot)$ in the non-positive real axis. Colormaps fulfilling the latter conditions are known as \textit{cyclic} \cite{kovesi2015good, mpl}. In addition, the resulting signal is upsampled by factor 2 using an interpolation filter \cite{harris2004multirate}.
An example outcome obtained using the present procedure is reported in \autoref{fig:phasesd-annni-i}.

The final step of the diagram construction consists of the plotting of the vector field in \eqref{eq:vfield-def}.
The approach makes use of the \textit{Runge–Kutta method} \cite{runge-kutta} and our reference implementation is part of the function
\verb|matplotlib.pyplot.streamplot| of the software package \textsc{Matplotlib} \cite{mpl}.
In the realm of differential equations, the Runge–Kutta method is a well-known algorithm for solving initial-value problems.
In our instance, the velocities on the lattice $\{\blambda_{i, j}\}$ are given by the gradient in \eqref{eq:vfield-def}.
The initial values instead, are the points on the boundary of the lattice.
Furthermore, a heuristic path of lattice points spiraling toward the center is added to the initial values, to improve the density of the streamlines.

\section{Solution of the order parameter discovery problem}
\label{section:ord-param-solution}
In this section, we expand on the solution of the order parameter discovery problem defined in \eqref{eq:ord-param-qcqp}.
Instead, the special case in \eqref{eq:ord-param-special} is treated in \myappendixref{section:ord-param-solution-special}.
We restate the problem for clarity and convenience, so
\begin{align}
    \nonumber
    \underset{M \in \hermset{m}}{\min}&
    \sum_{(i, j) \in I^+ \times I^-} \left(
    -\frac{\langle M \rangle_i^2}{p_i}
    +
    \frac{\langle M \rangle_j^2}{p_j}
    \right),\\
    \text{s.t.}\,& \|M\|_F^2 \le 1,
\end{align}
with $\langle M \rangle_i\coloneqq\traceop(\rho_0(\blambda_i)M)$ and $p_i\coloneqq \traceop(\rho_0(\blambda_i)^2)$.

We introduce the \textit{row-major vectorization operator} $\vecop(\cdot): \mathbb{C}^{n \times n} \to \mathbb{C}^{n^2}$ defined as
\begin{align}
    \vecop(M) \coloneqq \sum_{i=1}^n M\ket{i}\otimes \ket{i},
\end{align}
for any matrix $M$ of order $n$ in $\mathbb{C}$.
For matrices $A, B$ of order $n$ we will be using the identity
\begin{align}
    \label{eq:tr-vec-identity}
    \traceop\left(AB^\dagger\right) =& \vecop(A)^\dagger \vecop(B).
\end{align}
We recall that we denoted $\hermset{m}$ the set of Hermitians of order $m$ in $\mathbb{C}$.
We define the set of vectorized Hermitians of order $m$ as
\begin{align}
    \widehat{\hermset{m}} \coloneqq& \left\{\vecop(M)\middle| M \in \hermset{m}\right\}.
\end{align}
Let $\rho_i=\rho_0(\blambda_i)$, and let $I$ be some finite index set.
Assume $M \in \hermset{m}$, then we have that
\begin{subequations}
\begin{align}
    \label{eq:tr-squared}
    \sum_{i\in I} \frac{\langle M \rangle_i^2}{p_i} =& \sum_{i\in I}\frac{\left(\traceop(\rho_i M)\right)^2}{p_i}\\
    \underset{\eqref{eq:tr-vec-identity}}{=}&
    \sum_{i \in I}\frac{\left(\vecop(M)^{\dagger} \vecop(\rho_i)\right)^2}{p_i}\\
    =& \mathbf{x}^{\dagger} \left(\sum_{i \in I} \frac{\mathbf{r}_i\mathbf{r}_i^{\dagger}}{p_i} \right) \mathbf{x},
\end{align}
\end{subequations}
with $\mathbf{x}\coloneqq \vecop(M)$ and $\mathbf{r}_i\coloneqq\vecop(\rho_0(\blambda_i))$.
We note that the vectors $\mathbf{x}$ and $\mathbf{r}_i$ belong to the set of vectorized Hermitians $\widehat{\hermset{m}}$.

We use the result in \eqref{eq:tr-squared} to rewrite the optimization problem in \eqref{eq:ord-param-qcqp}
in the equivalent\footnote{Up to the positive constant (for the objective) $|I^+|\cdot |I^-|$.} form
\begin{subequations}
\begin{align}
    \label{eq:appx-ordp-opi}
    \underset{\mathbf{x} \in \mathbb{C}^{m^2}}{\min}\,& \mathbf{x}^{\dagger} A \mathbf{x},\\
    \label{eq:appx-ordp-opii}
    \text{s.t.}\,& \|\mathbf{x}\|^2_2 \le 1,\\
    \label{eq:appx-ordp-opiii}
    & \mathbf{x} \in \widehat{\hermset{m}},
\end{align}
\end{subequations}
with
\begin{align}
    \label{eq:ordp-mat-a-def}
    A \coloneqq& -\frac{1}{|I^+|}\sum_{i\in I^+} \frac{\mathbf{r}_i\mathbf{r}_i^{\dagger}}{p_i}
    +\frac{1}{|I^-|} \sum_{j\in I^-} \frac{\mathbf{r}_j\mathbf{r}_j^{\dagger}}{p_j}.
\end{align}
This formulation confirms that the optimization is a quadratically constrained quadratic program as stated in \autoref{section:ord-param}.
The matrix $A$ (determined by data) is Hermitian, so the objective in \eqref{eq:appx-ordp-opi} is real and well-defined (even in the absence of constraint \eqref{eq:appx-ordp-opiii}).
We expand on the condition for the non-degeneracy of the problem introduced in \autoref{section:ord-param}. We impose that the matrix $A$ is \textit{indefinite} \cite{horn},
where the Hermitian structure is fulfilled by the construction in \eqref{eq:ordp-mat-a-def}.
In other words, we require the matrix $A$ to have both (strictly) positive and negative eigenvalues\footnote{
    As a simple example, this condition is not met when the sets $I^{+}$ and $I^{-}$ are identical.
    However, this would be incoherent since there is no distinction between the phases.
}.

Let $(\cdot) \succcurlyeq (\cdot)$ denote the \textit{Loewner order} \cite{horn}, that is the partial order on the cone of PSD (positive semidefinite) matrices.
Specifically, for any pair of Hermitian matrices $X, Y$ we have that $X \succcurlyeq Y$ if and only if $X-Y$ is PSD.

The optimization problem is non-convex since we do not assume $A\succcurlyeq 0$, indeed $A$ is required indefinite.
However, as anticipated in \autoref{section:ord-param}, this optimization problem can be solved efficiently even in the case of the non-convexity of the objective (i.e., matrix $A$ is not PSD).
Moreover, this is an exceptional case where \textit{strong duality}\footnote{
    Strong duality is equivalent to the duality gap is zero, that is the difference between primal and dual solutions.
}~\cite{Boyd_Vandenberghe_2004} holds, provided that Slater's constraint qualification is fulfilled.
That is, there exists an $\mathbf{x} \in \widehat{\hermset{m}}$ such that the inequality constraint \eqref{eq:appx-ordp-opii} holds strictly (i.e. not tight).
In our case, an example is $\mathbf{x}=\vecop(\idenm{m})/\sqrt{m+\epsilon}$ for any $\epsilon>0$, so $\|x\|^2_2 < 1$.

Now, we consider the optimization problem consisting of \eqref{eq:appx-ordp-opi} and \eqref{eq:appx-ordp-opii}.
We exclude the constraint in \eqref{eq:appx-ordp-opiii}, as we will prove being enforced implicitly by the structure of the matrix $A$
and the non-degeneracy conditions.
The Lagrangian function is
$\mathcal{L}(\mathbf{x}, \alpha)=\mathbf{x}^{\dagger}(A+\alpha \idenm{m^2})\mathbf{x} - \alpha$ with the multiplier $\alpha \in \mathbb{R}_{+}$.
Consequently, the dual function $\underset{\mathbf{x}}{\min}\,\mathcal{L}(\mathbf{x}, \alpha)$ takes the value $-\alpha$ when $A+\alpha \idenm{m^2}\succcurlyeq 0$
(i.e. PSD, so the Lagrangian is bounded below in $\mathbf{x}$), and it becomes unbounded otherwise.
Let $\lambda_{\mathrm{min}}(\cdot)$ denote the minimum eigenvalue of the matrix argument.
The Lagrange dual problem reads
\begin{subequations}
\begin{align}
    \underset{\alpha \in \mathbb{R}_{+}}{\max}\,& -\alpha,\\
    \text{s.t.}\,& A+\alpha \idenm{m^2}\succcurlyeq 0,
\end{align}
\end{subequations}
then the constraint is fulfilled when $\alpha \ge -\lambda_{\mathrm{min}}(A)$, so the dual optimal is $\alpha^{\star}=-\lambda_{\mathrm{min}}(A)>0$.
The space of the primal solutions is given by the eigenspace corresponding to the minimum eigenvalue of $A-\lambda_{\mathrm{min}}(A)\idenm{m^2}$,
that is the null space of the latter.
Given any matrix $A$, we denote the null space of $A$ (i.e., the set of solutions of the homogeneous equation $Ax=0$) by $\mathrm{Null}(A)$.
Let
\begin{align}
    \label{eq:ordp-null-sp-sol}
    \mathbf{x}^{\star} \in \mathrm{Null}\left(A-\lambda_{\mathrm{min}}(A)\idenm{m^2}\right),
\end{align}
with $\|x^{\star}\|^2_2=1$, be an optimal solution\footnote{
    We note that the vectorization operator $\vecop(\cdot): \mathbb{C}^{m\times m} \to \mathbb{C}^{m^2}$ is an isomorphism, so the expression $\mathbf{x}^{\star}=\vecop(M)$
    implicitly means that the matrix $M$ is obtained uniquely from $\mathbf{x}^{\star}$ using the inverse of $\vecop$.
}, we show that if $\mathbf{x}^{\star}=\vecop(M)$, then $M=M^{\dagger}$. That is constraint \eqref{eq:appx-ordp-opiii} is implied by the structure of matrix $A$.

The set $\hermset{m}$ of Hermitian matrices of order $m$ is a real vector space, and if $\mathcal{B}$ is a basis for it,
then $\mathcal{B}_v\coloneqq\{\vecop(K)|K \in \mathcal{B}\}$ is a basis for the vector space of vectorized Hermitians.
We verify immediately that, by construction, the image of matrix $A$ belongs to the latter vector space.
By the non-degenerancy assumptions (i.e. $A$ is Hermitian indefinite), we have that $\lambda_{\mathrm{min}}(A) \ne 0$.
Consequently, the solution $\mathbf{x}^{\star}$ belongs to the image of $A$, hence the matrix $M$ such that $\vecop(M)=\mathbf{x}^{\star}$ is Hermitian.
In other words, we have shown that the assumption that $A$ is Hermitian indefinite implies that constraint \eqref{eq:appx-ordp-opiii} is fulfilled.
When the non-degeneracy conditions are not met, the case corresponds to the impossibility of 
obtaining an order parameter (which could be conditioned on the size of the observable). This is a situation we encountered in \autoref{section:experiments-ord-params}.

In \eqref{eq:ordp-null-sp-sol} we have proved that the solution may not be unique; however, this is consistent with the non-uniqueness of order parameters stated in \autoref{section:ord-param}.

We note that even in the case of 1-dimensional null space in \eqref{eq:ordp-null-sp-sol}, the optimal observable $M$ can be of any rank since the solution $\mathbf{x}^{\star}$ is a vectorization of a Hermitian operator. For example, in the case of the Ising model, we could have $\mathbf{x}^{\star}\approx \begin{pmatrix}1 & 0 & 0 & -1\end{pmatrix}^{\top}$ (a Bell's basis vector)
with the corresponding observable being $M=\sigma_z$ (full rank).

We summarize the procedure. Given a lattice of RDMs for the ground states of a selected Hamiltonian, we obtain the vector field in \eqref{eq:vfield-def}. The labeling of the phases is obtained using the trigonometric approach explained in \autoref{section:ord-param}, so we derive the sets $I^{+}$ and $I^{-}$. Subsequently to the instantiation of matrix $A$ given in \eqref{eq:ordp-mat-a-def}, we use its eigendecomposition to obtain, first, the verification that the non-degeneracy conditions are met. Secondly, the eigenspace corresponding to its minimum eigenvalue determines the space of solutions \eqref{eq:ordp-null-sp-sol}.

We conclude this section with a clarification of the contingency protocol integrated into our method.
We begin by assuming that a given QPT is identified via the RFS, enabling the construction of the phase diagram.
In accordance with our approach, data from RFS is employed to formulate the optimization problem for the corresponding order parameter.
In this study, we focus on linear order parameters, which may, in some cases, be insufficient to distinguish between phases (e.g., topological order).
However, the condition on the indefiniteness of the matrix in \eqref{eq:ordp-mat-a-def} ensures a fail-safe mechanism.
In other words, we can detect when the linear model is inadequate to capture the phase transition under investigation.

\newpart{We add the following clarification on the practical use of the indefiniteness condition.
In practice, we check that the matrix A (defined in Eq. \eqref{eq:ordp-mat-a-def}) has at least one numerically significant negative eigenvalue before proceeding with the optimization.
If A is positive semidefinite (all eigenvalues $\ge 0$), this signals that the current choice of subsystem size $k$ is insufficient to distinguish the phases under consideration, and we should increase $k$. This condition was encountered in the Rydberg model when attempting to distinguish $\mathbb{Z}_2$ from $\mathbb{Z}_3$ with $k = 1$ or $k = 3$, and was resolved by increasing to $k = 4$.}

\section{Special solution of the order parameter discovery problem}
\label{section:ord-param-solution-special}
In this section, we obtain the solution for the special case of the order parameter discovery established by the problem in \eqref{eq:ord-param-special}.
The outcome is the closed-form solution put forward in \eqref{eq:ord-param-special-em}.

Fixed the density matrices $\rhoa, \rhob$ of order $m$, we define the linear operator
$\Xi: \hermset{m} \to \hermset{m}$ on the space of Hermitian matrices, given by the rule
\begin{align}
	\label{eq:xi-op-def}
	\Xi(K) \coloneqq& -\frac{\left\langle K, \rhoa\right\rangle}{\left\langle \rhoa, \rhoa\right\rangle} \rhoa
	+ \frac{\left\langle K, \rhob\right\rangle}{\left\langle \rhob, \rhob\right\rangle} \rhob,
\end{align}
with $K \in \hermset{m}$.
Let $C \in \hermset{m}$ be any non-zero Hermitian, then the operator
\begin{align}
	\label{eq:gs-process-op}
	K \mapsto \frac{\left\langle K, C\right\rangle}{\left\langle C, C\right\rangle}C
\end{align}
is a projection onto the 1-dimensional subspace spanned by $C$.
Consequently, the operator in \eqref{eq:xi-op-def} can be related to the \textit{Gram-Schmidt process} (GS) \cite{golub}.
We obtain the quadratic form for the operator $\Xi$, so for $K \in \hermset{m}$
\begin{align}
	\label{eq:xi-qform}
	\langle K, \Xi(K) \rangle =& -\frac{\traceop\left(\rhoa K\right)^2}{\traceop(\rhoa^2)} +
	\frac{\traceop\left(\rhob K\right)^2}{\traceop(\rhob^2)}
	\in \mathbb{R}.
\end{align}
The latter is the objective of the problem in \eqref{eq:ord-param-special}, so the optimization can be rewritten as
\begin{align}
    \label{eq:xi-qform-optim}
    \underset{M \in \hermset{m}}{\min}&
    \langle M, \Xi(M) \rangle,\\
    \nonumber
    \text{s.t.}\,& \|M\|_F^2 \le 1.
\end{align}

We obtain the solution of the latter problem by means of the singular value decomposition (SVD) of the operator $\Xi$,
which is established by the following result.
\begin{theorem}
    \label{lemma:xi-op-svd}
	Let $\Xi: \hermset{n} \to \hermset{n}$ be the linear operator defined in \eqref{eq:xi-op-def}.
	Assume that for the density matrices $\rhoa, \rhob$, defining $\Xi$, it holds that $\langle\widehat{\rhoa}, \widehat{\rhob}\rangle\ne 1$.
	Then,
	\begin{align}
		\Xi(K) =& \|V_1\|_F \left\langle K, \widehat{V_1}\right\rangle U_1
		+ \|U_2\|_F \left\langle K, V_2\right\rangle \widehat{U_2}\\
		\label{eq:xi-op-svd}
		=&
		\sqrt{1-\langle \widehat{\rhoa}, \widehat{\rhob} \rangle^2} \cdot \left(
		\left\langle K, \widehat{V_1}\right\rangle U_1
		+ \left\langle K, V_2\right\rangle \widehat{U_2}
		\right)
	\end{align}
	is the reduced-SVD of $\Xi$, with non-zero singular values
    \begin{align}
        s_1=s_2=\sqrt{1-\langle \widehat{\rhoa}, \widehat{\rhob} \rangle^2},
    \end{align}
	and unnormalized right $V_i$ and left $U_i$ singular vectors
	\begin{subequations}
	\begin{align}
        \label{eq:xi-left-right-singular-vecs}
		V_1 =&
			\widehat{\rhoa}
			-\left\langle \widehat{\rhoa}, \widehat{\rhob} \right\rangle
			\widehat{\rhob};\quad
		U_1 = -\widehat{\rhoa}, \\
		V_2 =& \widehat{\rhob};\quad
		U_2 =
			\widehat{\rhob}
			-
			\left\langle \widehat{\rhoa}, \widehat{\rhob} \right\rangle
			\widehat{\rhoa}.
	\end{align}
	\end{subequations}
\end{theorem}
\begin{proof}
	We start by rewriting the operator $\Xi$ defined in \eqref{eq:xi-op-def} in the equivalent form
	\begin{align}
		\label{eq:xi-op-def-new}
		\Xi(K) =&
        - \left\langle K, \widehat{\rhoa}\right\rangle \widehat{\rhoa}
		+ \left\langle K, \widehat{\rhob}\right\rangle \widehat{\rhob}.
	\end{align}
	Consider the operator
	\begin{align}
		P(K)=&\left\langle K, \widehat{\rhob}\right\rangle\,
		\left\langle \widehat{\rhob}, \widehat{\rhoa}\right\rangle\,
		\widehat{\rhoa},
	\end{align}
	then adding and subtracting the latter to $\Xi$ we obtain
    \begin{subequations}
	\begin{align}
		\label{eq:gamma-gs}
		\Xi(K) =&
        \left(- \left\langle K, \widehat{\rhoa}\right\rangle +
        \left\langle K, \widehat{\rhob}\right\rangle\,
		\left\langle \widehat{\rhob}, \widehat{\rhoa}\right\rangle
        \right)\cdot \widehat{\rhoa}\\
        \nonumber
		&+ \left\langle K, \widehat{\rhob}\right\rangle \cdot \left(\widehat{\rhob} - 
		\left\langle \widehat{\rhob}, \widehat{\rhoa}\right\rangle\, \widehat{\rhoa}\right)\\
        \nonumber
        =&
        \left\langle K, \widehat{\rhoa} - \left\langle \widehat{\rhob}, \widehat{\rhoa}\right\rangle \widehat{\rhob}  \right\rangle
        \cdot \left(-\widehat{\rhoa}\right)
		+ \left\langle K, V_2\right\rangle U_2\\
        \nonumber
        =& \left\langle K, V_1\right\rangle U_1 
		+ \left\langle K, V_2\right\rangle U_2,
	\end{align}
    \end{subequations}
    with $V_i$ and $U_i$ defined in \eqref{eq:xi-left-right-singular-vecs}.
	In general, we have that $\langle \rhoa, \rhob\rangle \ne 0$, so $K$ can have a non-zero projection on both terms of \eqref{eq:xi-op-def-new}.
	However, in \eqref{eq:gamma-gs} we can immediately verify that $\langle V_1, V_2 \rangle=0$ and $\langle U_1, U_2 \rangle=0$.
	In addition, we verify that
	\begin{align}
		\langle V_1, V_1 \rangle=& \langle U_2, U_2 \rangle = 1-\langle \widehat{\rhoa}, \widehat{\rhob} \rangle^2 \in (0, 1],
	\end{align}
	thus we can rewrite \eqref{eq:gamma-gs} as stated in \eqref{eq:xi-op-svd}.
    Hence, the expression in \eqref{eq:xi-op-svd} is the reduced-SVD of $\Xi$, as stated in the claim.
\end{proof}

As the final step, we identify the close formulation for the solution of the problem in \eqref{eq:xi-qform-optim}, using the SVD of the operator $\Xi$.
We assume the conditions of \autoref{lemma:xi-op-svd}.
Then, by substituting the right singular vectors into the quadratic form in \eqref{eq:xi-qform} we obtain
\begin{subequations}
\begin{align}
    \left\langle \widehat{V_1}, \Xi(\widehat{V_1}) \right\rangle =&
    1 - \frac{\traceop\left(\rhoa \rhob\right)^2}{\traceop(\rhoa^2)} > 0,\\
    \left\langle \widehat{V_2}, \Xi(\widehat{V_2}) \right\rangle =&
    -\left(1-\langle \widehat{\rhoa}, \widehat{\rhob} \rangle^2\right) < 0.
\end{align}
\end{subequations}
Hence $M=\widehat{V_2}$ (expanded in \eqref{eq:ord-param-special-em}) is the observable that minimizes the objective in \eqref{eq:xi-qform},
subject to the norm of the Hermitian $M$ being 1.
In \myappendixref{section:exact-ordp-ising} we showcase the exact formula applied to the Ising model.

\section{Exact example for the order parameter of the Ising model}
\label{section:exact-ordp-ising}
We consider the quantum Ising model in a transverse magnetic field 
and the representative density matrices $\rhoa, \rhob$ for its phases: if the interaction term dominates or the magnetic field strength dominates, respectively. We denote with $\sigma^x, \sigma^y, \sigma^z$ the corresponding Pauli matrices.
Let $\rhoa=\frac{\idenmnodim + \sigma^z}{2}$ and $\rhob=\frac{\idenmnodim + \sigma^x}{2}$. Both $\rhoa$ and $\rhob$ are norm 1 orthogonal projections,
and $\langle \rhoa, \rhob\rangle=\frac{1}{2}$.
We apply the formula for the optimal observable in \eqref{eq:ord-param-special-em} to get
\begin{align}
	\label{eq:k-star-ising}
	M^{\star}=& \frac{\idenmnodim + 2\sigma^z - \sigma^x}{2\sqrt{3}}
    = \frac{\sigma^z}{\sqrt{3}} + \frac{\idenmnodim- \sigma^x}{2\sqrt{3}}
\end{align}
which is a norm 1 Hermitian. Clearly, for the disordered phase $\traceop(\rhob M^{\star})=0$, whereas $\traceop(\rhoa M^{\star}) = \frac{\sqrt{3}}{2}$ for the ordered phase. Moreover, we note that $M^{\star}$ is a linear combination of the orthogonal components $\sigma^z$ and $I-\sigma^x$. The $\sigma^z$ operator is the local order parameter or scaling operator, while $I-\sigma^x$ measures the magnetization in the $x$-direction.

Now, let $M=\sigma^z/\|\sigma^z\|_F=\sigma^z/\sqrt{2}$ (norm 1 as for $M^{\star}$), that is the observable $M$ is the magnetization in the $z$-direction for present model. Then, $\traceop(\rhob M)=0$ and $\traceop(\rhoa M)=\frac{1}{\sqrt{2}}$, hence we confirm that
\begin{align}
    \traceop(\rhoa M) \le \traceop(\rhoa M^{\star}).
\end{align}
In other words, we have shown that the observable $M^{\star}$ presents an increased gap between the phases, when compared to the known magnetization $M$, or it maximizes the value in the order phase.

\section{Experimental Methods}
\label{section:dmrg_computation}
To calculate the ground states for the reduced fidelity susceptibility, we use the density matrix renormalization algorithm (DMRG) \cite{Schollw_ck_2011}.
We use two different software packages called Tensor Network Python \textsc{TeNPy} \cite{tenpy} and Quantum Simulation with MPS Tensor \textsc{qs-mps} \cite{qs-mps}. 
In the case of the Rydberg model instead, we use \textsc{ITensor} \cite{Fishman_2022}.
  
For the ANNNI model we span the region $(\kappa, h) \in \mathcal{R}=[0.01, 1.5] \times [0.01, 1.5]$ while for the cluster $(K, h) \in \mathcal{R}=[0.5, 1.5] \times [0.5, 1.5]$. We take $n=64$ points for each axis, resulting in a $64 \times 64$ lattice of parameters, with a maximum bond dimension $\chi = 64$. Both Hamiltonians in the top-left corner of their respective regions $\mathcal{R}$ are dominated by the $\sigma^z$ term and thus the states will be a perturbation of the all-up state $\ket{\uparrow \uparrow \cdots \uparrow}$. 
\newpart{To ensure the convergence of the DMRG calculations we use the following strategy: we begin from a product state corresponding to a simple, trivial phase, and then progressively vary the couplings, using the previously converged ground state as the initial guess for the next computation. To make the procedure more robust near phase transitions, where convergence may become difficult, we implement a backup strategy. In these regions, the ground state obtained in the previous iteration may no longer provide a good initial guess. Therefore, whenever the computation time exceeds a certain threshold, defined from the average runtime of previous calculations, we restart the algorithm from a random state, which may have a larger overlap with the true ground state close to the phase transition. We note the most time-consuming part of the calculation is solving the local eigenvalue problem through the method} \verb|eigsh| 
\newpart{from the software package \textsc{SciPy} \cite{SciPy}. This procedure is effective up until the point of a phase transition. At that point, if the computation time exceeds a specified threshold (which can be set either arbitrarily or adaptively), we shelve the calculation and proceed to the next lattice point.}

\newpart{We perform a comparison between global and reduced fidelity susceptibility. In the former case, we take the fidelity between the MPS representing the ground state $\ket{\psi_0 (\blambda)}$ for the Hamiltonian parameters $\blambda =  (\lambda_1,\lambda_2)$, and $\ket{\psi_0 (\blambda + \bdelta)}$ for $(\lambda_1 +\delta,\lambda_2)$, $\bdelta = (\delta,0)$. In the case of the RFS, we compute the Uhlmann fidelity \eqref{eq:fun-f-def} between RDMs of $k$ sites starting from the ground states above mentioned.}

\begin{figure}[h!]
    \centering
    \includegraphics[width=1.1\linewidth]{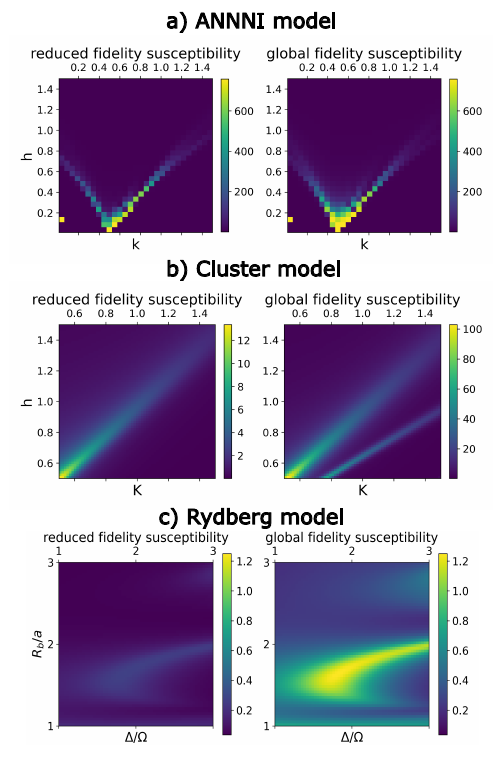}
    \caption{Comparison of global \emph{vs} reduced fidelity susceptibility. \textbf{a)} For the ANNNI model, we use a chain of $L=20$ and a grid of $(n \times n), \quad n=30$ points\newpart{, and a two-site RDM}. They show the same features of ising-like transition and the incommensurate-commensurate PT transition. \textbf{b)} For the Cluster, model we use a chain of $L=50$ and a grid of $(n \times n), \quad n=64$ points. Both methods capture the symmetry-protected phase transition, but we notice that an artifact is present in the case of global susceptibility. This is probably due to a longitudinal term $\epsilon= 1e-3$ we have in every tensor. \textbf{c)} For the Rydberg model, we use the \newnewpart{chain length $L=30$ and a parameter grid of $64 \times 64$. We use a four-site RDM so that we can capture both crystalline lobes.}}
    \label{fig:comparison-gfs-rfs}
\end{figure}

\newpart{Then, after obtaining the global and reduced fidelities, we can use the second discrete derivative as described in \eqref{eq:second-discr-der}. To make the comparison fair, we do not show the vector field in \eqref{eq:vfield-def} or the angular scalar function in \eqref{eq:angle-grad} as in \autoref{fig:phasesd-annni-i}. Hence, we show in \autoref{fig:comparison-gfs-rfs} the three models studied in our manuscript: ANNNI, Cluster, and Rydberg.}
The figure shows in general that the two methods give comparable results.
\newnewpart{The most noticeable difference between these plots and the ones in the main text is that our method relates the fidelity susceptibility to a geometric information tool giving us clearer phase boundaries and knowledge of the sources and sinks of the vector field.}

\section{The ANNNI model}
\label{sec:annnitheory}
We first present the phase diagram for the ANNNI Model, which is given in \autoref{fig:annni-theory}, where a variety of rich phase transitions are present. We initially begin with small $\kappa$ and $h$, where the interactions between neighbors along the x-axis dominate, resulting in a ferromagnetic phase where all spins align parallel to one another, either as $\ket{\rightarrow \rightarrow \cdots \rightarrow}$ or $\ket{\leftarrow \leftarrow \cdots \leftarrow}$. This ferromagnetic region is highlighted in purple in \autoref{fig:annni-theory}.

Upon increasing $\kappa$ and $h$, a phase transition occurs as the spins enter a paramagnetic phase (PM), presented in grey in  \autoref{fig:annni-theory}. Entering the paramagnetic phase, the transverse magnetic field begins to dominate, causing all states to align with the magnetic field, resulting in the state $\ket{\uparrow \uparrow \cdots \uparrow}$. This Ising-like transition has been previously studied in \cite{Suzuki2013} and the transition line (which corresponds to the purple line in \autoref{fig:annni-theory} is 
given in \cite{cea2024exploring, Suzuki2013} as;
\begin{equation}
    h_I \approx \frac{1 - \kappa}{\kappa} \left( 1 - \sqrt{\frac{1 - 3\kappa + 4\kappa^2}{1 - \kappa}} \right).
\end{equation}

Increasing the parameters $h$ and $\kappa$, a further transition becomes evident in the phase diagram between the paramagnetic and floating phases (FP). This incommensurate-incommensurate  Kosterlitz-Thouless (KT) phase transition \cite{PhysRevLett.42.65, PhysRevB.29.239, 1982RPPh...45..587B} (given in blue in \autoref{fig:annni-theory}) has been approximated in \cite{PhysRevB.76.094410} to be
\begin{equation}
    h_{KT}(\kappa) \approx 1.05 \sqrt{\left( \kappa - \frac{1}{2} \right) (\kappa - 0.1)}.
\end{equation}

Such a transition corresponds to a change in the modulation wave vector \cite{2019arXiv190309179V}, leading to different incommensurate structures along the spin chain.

A further transition evident in this phase diagram is the transition from the floating phase to the antiphase (AP), where the states exhibit staggered magnetization, as a result of the next-nearest neighbor interactions dominating, corresponding to $\ket{\cdots\leftarrow \leftarrow \rightarrow \rightarrow \leftarrow \leftarrow\cdots}$. Such an incommensurate-commensurate transition (given in green in  \autoref{fig:annni-theory}) is described by the Pokrovsky-Talapov universality class \cite{PhysRevLett.42.65, PhysRevB.66.024412}, with the transition in this case given by 
\begin{equation}
h_{PT} \approx 1.05(\kappa - \frac{1}{2})
\end{equation}
as highlighted in \cite{PhysRevB.76.094410} and corresponds to the green line in \autoref{fig:annni-theory}.
We also note the presence of the Lifshitz point, marked by a red dot in  \autoref{fig:annni-theory}. This point represents a critical point, known as the Lifschitz point (LP), on the phase diagram of certain magnetic systems, where a line of second-order phase transitions meets a line of incommensurate modulated phases \cite{article}. Finally, we note the presence of a disordered line in the ANNNI model, referred to as the Peschel-Every (PE) line \cite{Peschel1981-zp} within the paramagnetic phase. This line serves as a reference point for understanding the characteristics of this specific phase.

We note this model also reproduces important features observed experimentally in systems that can be described by discrete models with effectively short-range competing interactions \cite{Selke1988TheAM}. These experimental findings include Lifshitz points \cite{Henkel2010, HORNREICH1980387}, adsorbates, ferroelectrics, magnetic systems, and alloys. Conversely, the so-called floating phase emerging in the model is appealing to experimental researchers to explore. This critical incommensurate phase has been observed very recently by using Rydberg-atom ladder arrays \cite{Zhang:2024utj}.

\section{Vector field patterns}
\label{section:vfield-patterns}

\begin{figure}[h]
    \centering
    \begin{tikzpicture}[scale=1.5]
\draw[->,thick, black!60] (-0.1,0)--(4.5,0) node[right]{$x_1$};
\draw[->,thick, black!60] (0,-0.75)--(0,2.0) node[above]{$x_2$};

\draw[domain=0:4, smooth, variable=\t, black, thick] plot ({\t}, {sin(\t r)}) node[right] {$\mathbf{c}(t)$};
\pgfmathsetmacro{\tzero}{2}
\pgfmathsetmacro{\xzero}{\tzero}
\pgfmathsetmacro{\yzero}{sin(\tzero r)}

\foreach \t in {0.3, 0.5, 0.7, 0.9, 1.1, 1.3, 1.5, 1.7, 3.1, 3.3, 3.5, 3.7, 3.9}
{
    \pgfmathsetmacro{\x}{\t}
    \pgfmathsetmacro{\y}{sin(\t r)}
    \pgfmathsetmacro{\tangentx}{0.5 * 1}
    \pgfmathsetmacro{\tangenty}{0.5 * cos(\t r)}
    \pgfmathsetmacro{\normalx}{-\tangenty}
    \pgfmathsetmacro{\normaly}{\tangentx}
    \draw[black,->, densely dashed] (\x, \y) -- ++(\normalx, \normaly);
}

\filldraw[black] (\xzero, \yzero) circle (1pt) node[above=2pt, right=2pt] {$\mathbf{c}(t_0)$};

\pgfmathsetmacro{\tangentx}{1}
\pgfmathsetmacro{\tangenty}{cos(\tzero r)}
\draw[black,->,thick, densely dotted] (\xzero, \yzero) -- ++(\tangentx, \tangenty) node[above right] {$\mathbf{c}^{\prime}(t_0)$};

\pgfmathsetmacro{\normalx}{-\tangenty}
\pgfmathsetmacro{\normaly}{\tangentx}
\draw[black,->,thick, densely dotted] (\xzero, \yzero) -- ++(\normalx, \normaly) node[above right] {$\nabla_\mathbf{x} g(\mathbf{c}(t_0))$};
\end{tikzpicture}
    \caption{Curve $\mathbf{c}(t)$ whose image belongs to a level set of a function $g: \mathcal{X} \subseteq \mathbb{R}^2 \to \mathbb{R}$, that is $g(\mathbf{c}(t))$ is constant
    for $t$ in some open interval.}
    \label{fig:sources-theory}
\end{figure}

This appendix aims to highlight the patterns that emerge in the plot of the vector fields. Specifically, we focus on the "source"-like patterns observed at the phase transitions and the "sink"-like patterns that become evident. We demonstrate the conditions under which these patterns appear in the plot of the vector field \( P(\blambda) \) (defined in \eqref{eq:vfield-def}).

Let $g:\mathcal{X} \subseteq \mathbb{R}^n \to \mathbb{R}$ be a differentiable function and define the \textit{level set} \cite{apostol1969calculus}
\begin{align}
    L(k) \coloneqq& \left\{\mathbf{x}\middle| \mathbf{x} \in \mathcal{X}, g(\mathbf{x})=k\right\}
\end{align}
for some $k\in \mathbb{R}$, as the subset of the domain of $g$ whose image under $g$ is the constant $k$.

Consider a differentiable curve $\mathbf{c}(t)$ on the level set $L(k)$, that is $\mathbf{c}: I \to L(k)$ for some open interval $I \subseteq \mathbb{R}$.
Consequently, we have that $g(\mathbf{c}(t))=k$ for all $t \in I$.
So, by the chain rule, we obtain
\begin{align}
    \label{eq:norm-vs-tangent-curve}
    \left\langle\mathbf{c}^{\prime}(t), \nabla_\mathbf{x} g(\mathbf{c}(t))\right\rangle =0,
\end{align}
where $\langle\cdot, \cdot\rangle$ is the inner product on $\mathbb{R}^n$.
In other words, if the vector arguments in \eqref{eq:norm-vs-tangent-curve} are non-zero, then they must be orthogonal.
So, the non-zero gradient of $g$ at a point $\mathbf{c}(t)$ is perpendicular to the tangent of the curve $\mathbf{c}$ at the same point.
A pictorial example of the case is given in \autoref{fig:sources-theory}.

We recall that the gradient is the direction of the greatest rate of increase of a function.
Then, the source-like structures we observe on the plot of the vector field $P(\blambda)=-\nabla g(\blambda)$ (eg. see \autoref{fig:cluster}) are the local maxima of the fidelity susceptibility $g$ in the direction transversal to the phase transition. The latter is consistent with the manifestation of QPT as a maxima of the susceptibility in finite-size systems.

\section{Scale invariance of the principal argument}
\label{section:scale-invariant}
We prove the scale-invariant property of the function $\theta(\blambda)$ defined in \eqref{eq:angle-grad},
which is employed in the construction of the phase diagrams.

We begin with a few concepts on the class of homogeneous functions.
We call a function $f: K \subseteq \mathbb{R}^n \to \mathbb{R}$ \textit{positive homogeneous} (PH) of degree $\alpha$ when
\begin{align}
    \label{eq:homog-fun-def}
    f(t \mathbf{x})=t^{\alpha} f(\mathbf{x}) \propto f(\mathbf{x})
\end{align}
for all scaling factors $t > 0$ such that $t x \in K$.
The term "positive" refers to the restriction on the scaling factor $t>0$, so the degree $\alpha$ can be any real number.
The \textit{power law} $x \mapsto c|x|^{\alpha}$ is PH.
The parameter $\alpha$ is known as \textit{critical exponent} in physics.
Homogeneity is preserved by differentiation, so for $f: \mathbb{R} \to \mathbb{R}$
\begin{align}
    \frac{\partial f(t x)}{\partial x} =& t f^{\prime}(t x) = t^{\alpha} f^{\prime}(x),
\end{align}
(the right-most equality follows from the definition in \eqref{eq:homog-fun-def})
dividing by $t$ we get $f^{\prime}(t x) = t^{\alpha - 1} f^{\prime}(x)$ so $f^\prime$ is PH with degree $\alpha  - 1$.
The latter shows that the vector field in \eqref{eq:vfield-def} preserves PH.
Instead, the angle of the vector field defined in \eqref{eq:angle-grad} is scale-invariant (when the susceptibility follows the power law as we approach QPT).
Indeed, by denoting the partial derivative $g_{\lambda_i}=\partial g/\partial \lambda_i$ of the function $g$ defined in \eqref{eq:fun-g-def}, we have that
\begin{align}
    \theta((\blambda - \blambda_c) t) =& \arctan\left(
        \frac{
            t^{\alpha - 1} g_{x_2}(\blambda - \blambda_c)
        }{
            t^{\alpha - 1} g_{x_1}(\blambda - \blambda_c)
        }
    \right) = \theta(\blambda - \blambda_c),
\end{align}
when $g_{x_1}(\blambda - \blambda_c)>0$ (the other cases for $\mathrm{atan2}(\cdot, \cdot)$ follow similarly),
with $\blambda_c$ denoting the critical value for the QPT.

\section{Relation to information geometry}
\label{section:info-geo-link}
In this appendix, we discuss the fidelity susceptibilities relationship to information geometry and the \textit{Quantum Fisher Information} (QFI). 
In \eqref{eq:sq-bures-dist-def} we introduced the Bures distance as the statistical distance between density matrices.
The latter, when considered in terms of a small displacement $\mathrm{d}\blambda$ in the vector of parameters $\blambda$, induces the \textit{Bures metric} \cite{paris-qest}
\begin{align}
    d^2_B(\rho(\blambda), \rho(\blambda + \mathrm{d}\blambda)) =&
    \sum_{i, j} h^{i, j}_{\blambda} \mathrm{d}\lambda_i\,\mathrm{d}\lambda_j,
\end{align}
where $h_{\blambda}$ is the \textit{metric tensor} at $\blambda$.
In other words, the Bures metric is a notion of distinguishability of two density matrices that are infinitesimally close in the coordinate system $\blambda$.
In the realm of condensed matter physics, the Bures metric is better known as the fidelity susceptibility (introduced in \autoref{sec:prelim}).
Thus, the singularities of the metric $h_{\blambda}$, correspond to the critical regions of the underlying model.

In quantum metrology, the \textit{quantum Cram\'er-Rao bound} \cite{helstom-qdet-est, PhysRevLett.72.3439}, is the quantum equivalent to the classical lower-bound \cite{Rao1992}
of the variance of the unbiased estimator $\widehat{\lambda}$ (i.e. $\mathbb{E}[\widehat{\lambda}]=\lambda$) for an unknown parameter $\lambda$.
The bound, in the case of a single parameter $\lambda$, is given by
\begin{align}
    \mathrm{Var}(\widehat{\lambda}) \ge \frac{1}{N Q(\rho(\lambda))},
\end{align}
where $N$ is the number of independent repetitions and $Q(\rho(\lambda))$ is the QFI \cite{paris-qest}.
We consider a multi-parameter estimator $\widehat{\blambda}$ the bound of the covariance matrix becomes
\begin{align}
    \mathrm{Cov}(\widehat{\blambda}) \succcurlyeq \frac{Q(\rho(\blambda))^{-1}}{N},
\end{align}
where the relation $(\cdot) \succcurlyeq (\cdot)$ is the Loewner order
(i.e. for $n\times n$ Hermitian matrices $A, B$, we have that $A\succcurlyeq B$ if and only if $A-B$ is PSD).

There was consensus that, in the general case, the fidelity susceptibility is proportional to the quantum Fisher information \cite{liu-fidelity-sus-qfi}, that is
\begin{align}
    \mathcal{X}_F(\lambda) \overset{?}{=}& \frac{1}{4} Q(\rho(\lambda)).
\end{align}
However, in \cite{PhysRevA.95.052320} it is proved that at the points where we have a rank change of the density matrix $\rho(\lambda)$, the two quantities are not the same
and the QFI presents a discontinuity.
We note that the meaning and the consequent violation of the Cram\'er-Rao theorem on such singularities is still an active area of research \cite{Seveso_2020, PhysRevA.100.032317}.

Concerning our formulation, the function $g(\blambda)$ defined in \eqref{eq:fun-g-def}, can be related directly to the trace of the metric tensor,
that is $g(\blambda)=-\traceop(h_{\blambda})$.
The minus factor is used as a convention to make the phase transitions appear as sources in the vector field resulting from \eqref{eq:vfield-def}.
In regions of the parameter space $\mathcal{X}$ where the rank of $\rho(\blambda)$ is constant, it holds that $g(\blambda)$ is proportional to the trace of the QFI matrix.
We also note that the trace of $h_{\blambda}$ corresponds to the sum of the susceptibilities for each coupling parameter $\lambda_i$.
The choice of the trace is also justified by the fact that this operator is basis-independent.

\newpart{The function $g(\blambda)$ is the trace of the Bures metric tensor of the RDM manifold, but distinct from the global fidelity susceptibility computed from the full state. Our $g(\blambda)$ uses only the RDM, making it more accessible experimentally and in DMRG. Our formulation adds up contributions from all parameter directions $\blambda$, weighted by the Laplacian, and constructs a vector field from the result.}

For a comprehensive introduction to information geometry and QPT, we refer to \cite{PhysRevLett.99.100603}.

\clearpage
\pagebreak
\newpage

\bibliographystyle{quantum}
\bibliography{refs}
\end{document}